\documentclass[notitlepage,pra,amsmath,amssymb,onecolumn,superscriptaddress]{revtex4-1}
\usepackage{graphicx}
\usepackage{dcolumn}
\usepackage{bm}
\usepackage{color}

\begin{document}
\title{Determination of the Position of a Single Nuclear Spin from Free Nuclear Precessions Detected by a Solid-State Quantum Sensor}
\author{Kento Sasaki}
\affiliation{\mbox{School of Fundamental Science and Technology, Keio University, 3-14-1 Hiyoshi, Kohoku-ku, Yokohama 223-8522, Japan}}
\author{Kohei M. Itoh}
\email{kitoh@appi.keio.ac.jp}
\affiliation{\mbox{School of Fundamental Science and Technology, Keio University, 3-14-1 Hiyoshi, Kohoku-ku, Yokohama 223-8522, Japan}}
\affiliation{Spintronics Research Center, Keio University, 3-14-1 Hiyoshi, Kohoku-ku, Yokohama 223-8522, Japan}
\author{Eisuke Abe}
\email{e-abe@keio.jp}
\affiliation{Spintronics Research Center, Keio University, 3-14-1 Hiyoshi, Kohoku-ku, Yokohama 223-8522, Japan}
\date{\today}

\begin{abstract}
We report on a nanoscale quantum-sensing protocol which tracks a free precession of a single nuclear spin and
is capable of estimating an azimuthal angle---a parameter which standard multipulse protocols cannot determine---of the target nucleus.
Our protocol combines pulsed dynamic nuclear polarization, a phase-controlled radiofrequency pulse, and a multipulse AC sensing sequence with a modified readout pulse.
Using a single nitrogen-vacancy center as a solid-state quantum sensor, we experimentally demonstrate this protocol on a single $^{13}$C nuclear spin in diamond
and uniquely determine the lattice site of the target nucleus.
Our result paves the way for magnetic resonance imaging at the single-molecular level.
\end{abstract}
\maketitle

Nuclear magnetic resonance (NMR) spectroscopy is an analytical technique extensively used in chemistry, biology, and medicine.
It achieves sub-ppm spectral resolutions to provide a wealth of information on the structure and chemical environment of molecules,
but requires at least nanoliter-volume analytes containing an ensemble of identical nuclei, due to the insensitive induction detection the technique relies on.
In recent years, single isolated electron spins in solids, most prominently those associated with nitrogen-vacancy (NV) centers in diamond~\cite{SCLD14,RTH+14,AS18},
have emerged as atomic-scale quantum sensors
capable of detecting weakly-coupled external nuclei in as small as zeptoliter volumes: a dramatic decrease compared with conventional NMR~\cite{MKS+13,SSP+13,LPMD14}.
Furthermore, various NMR protocols, in which a train of properly-timed microwave pulses interrogates precessing nuclear spins via the interaction with the sensor electron spin,
have been devised and applied to external nuclei, demonstrating identifications of isotopes~\cite{MKC+14,HSR+15,DPL+15},
detection of single protons~\cite{SLC+14},
spectroscopy of single proteins~\cite{LSU+16},
spectral resolution approaching that of conventional NMR~\cite{APN+17,GBL+18},
and so on~\cite{SRP+15,KSD+15,PHHH16,KJM+17}.
A far-reaching yet natural goal of this line of research is chemical structure analysis at the single-molecular level,
{\it i.e.,} the determination of chemical identities and locations of the constituent nuclei in a single molecule.

\begin{figure*}
\begin{center}
\includegraphics{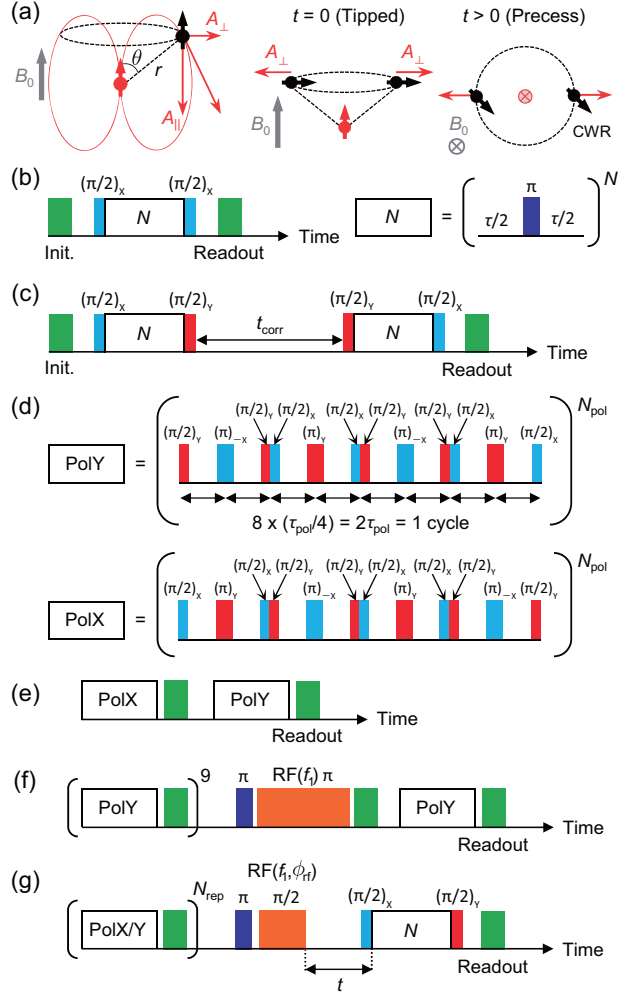}
\caption{\label{fig1}
(a) Schematic showing the sensor electron spin (in red) and the target nuclear spin (in black) coupled via the dipolar interaction under the static magnetic field $B_0$ (left).
When a nuclear spin is tipped, the direction of the hyperfine field relative to the nuclear spin is dependent on the azimuthal angle (middle),
and this information is reflected on the phase of a subsequent nuclear precession (right).
CWR: clockwise rotation.
(b--g) Pulse sequences used in the present work.
(b) Carr-Purcell (CP) sequence.
(c) Correlation spectroscopy.
(d) PulsePol sequences for DNP.
(e) Successive application of PolX and PolY to examine the polarization transfer.
(f) Sequence to selectively polarize a single nuclear spin.
(g) Protocol to determine the azimuthal angle.
}
\end{center}
\end{figure*}

For nuclei dipolarly-coupled with a sensor, the knowledge of the hyperfine parameters $A_{\parallel}$ and $A_{\perp}$ (the parallel and perpendicular components, respectively)
translates to the coordinate parameters $r$ and $\theta$, the distance from the sensor and the tilt (polar angle) from the applied static magnetic field $\bm{B}_0$, respectively,
owing to the form of the interaction $\propto (3 \cos^2 \theta - 1)/r^3$ or $3 \cos \theta \sin \theta/r^3$ [Fig.~\ref{fig1}(a, left)].
$^{13}$C nuclei ($I$ = $\frac{1}{2}$) in diamond sensed by the NV electronic spin ($S$ = 1) have served
as a canonical testbed for various NMR protocols to characterize the hyperfine parameters~\cite{AS18,ZHH+11,LHRM11,KUBL12,TWS+12,ZHS+12,LDB+13,BCA+16,RZB+17}.
For instance, in correlation spectroscopy, a multipulse sequence is repeated with the interval of $t_{\mathrm{corr}}$, during which a target nuclear spin evolves freely.
Boss {\it et al.} demonstrated that, by engineering the Hamiltonian during the free nuclear precession, both $A_{\parallel}$ and $A_{\perp}$ can be estimated with high precisions~\cite{BCA+16}.
The information that is still missing in order to {\it uniquely} determine the position of a target nuclear spin is the azimuthal angle $\phi$~\cite{LPM15,WHCP16,WCP17},
which, due to the symmetry of the interaction, does not formally appear in the Hamiltonian.

In this work, we show that the azimuthal angle can be determined by a multipulse protocol combined with dynamic nuclear polarization (DNP) and radiofrequency (RF) control of a target spin.
As a proof of principle, we apply this protocol to the NV--$^{13}$C -coupled system,
and uniquely assign the lattice site that the $^{13}$C nucleus sits in, even when multiple sites equivalent to it exist.

Our protocol proceeds as follows [see Fig.~\ref{fig1}(b--g) for the pulse sequences used].
The target nuclear spin is first polarized by transferring the polarization of the sensor electron spin using a multipulse sequence called PulsePol~\cite{SST+17},
and is tipped to the $xy$ plane by an RF $\pi$/2 pulse.
The dipolar field experienced by this tipped nuclear spin is dependent on $\phi$;
an example to appreciate this is that, when two nuclear spins are located on opposite sides of the sensor, the directions of the dipolar fields are also opposite [Fig.~\ref{fig1}(a, middle)].
A subsequent free precession of each nuclear spin carries this information as the initial phase of the oscillation [Fig.~\ref{fig1}(a, right)].

A central issue that must be addressed in this scenario is how to detect the phase of a nuclear precession.
Here, we consider a particular example in which $|m_S = 0 \rangle$ and $|m_S = -1 \rangle$ sublevels of the NV spin serve respectively
as $|0 \rangle$ and $|1 \rangle$ of the sensor two-level system, but the concept is general.
To detect a dynamics of a single nuclear spin,
the sensor spin is prepared in a superposition of $|m_S = 0 \rangle$ and $|m_S = -1 \rangle$,
and is subject to a Carr-Purcell (CP) sequence~\cite{CP54} consisting of $N$ $\pi$ pulses equally separated by $\tau$ [Fig.~\ref{fig1}(b)].
Because only $|m_S = -1 \rangle$ can hyperfine-couple with the nuclear spin,
the two sensor states act differently on the nuclear spin, making it rotate around different axes, say $\bm{n}_0$ and $\bm{n}_1$.
Different nuclear dynamics are combined by the final ($\pi$/2)$_{\mathrm{X}}$ pulse ($\pi$/2 pulse with the X phase defined in the rotating frame of reference).
The transition probability $P_{\mathrm{X}}$ of the sensor is given by~\cite{TWS+12}
\begin{equation}
P_{\mathrm{X}} = 1 -\frac{1}{2} (1 - \bm{n}_0 \cdot \bm{n}_1) \sin^2 \frac{N \phi_{\mathrm{cp}}}{2}.
\label{PX}
\end{equation}
When $\tau$ is chosen properly, $\phi_{\mathrm{cp}}$ and $\bm{n}_0 \cdot \bm{n}_1$ carry the information on a single nuclear spin.
This readout pulse has been extensively used to probe the nuclear dynamics in the past~\cite{TWS+12,TCS+14}, but does not depend on the free precession angle of the nuclear spin.

In our protocol, we instead use a ($\pi$/2)$_{\mathrm{Y}}$ pulse for readout. The transition probability becomes
\begin{equation}
P_{\mathrm{Y}} = \frac{1}{2} - \frac{1}{2} \cos (\phi - \phi_{\mathrm{n}}) \sin N \phi_{\mathrm{cp}},
\label{PY}
\end{equation}
where $\phi_{\mathrm{n}}$ is the azimuthal angle of the nuclear spin measured in real space.
A detail deviation of Eq.~(\ref{PY}) is given in Supplemental Material~\cite{SM}.
We measure a free precession of a nuclear spin, which oscillates as $\cos (2 \pi f_{\mathrm{p}} t + \phi_0)$.
$\phi$ is determined as $\phi = \phi_{\mathrm{n}}(0) + \phi_0$.
$\phi_{\mathrm{n}}(0)$ is the azimuthal angle of the nuclear spin at $t$ = 0 [defined as the end time of an RF pulse, Fig~\ref{fig1}(g)].
While to our knowledge Eq.~(\ref{PY}) has not been explicitly given in previous literature,
a recent demonstration of high-resolution spectroscopy of nuclear spin ensembles does employ this ($\pi$/2)$_{\mathrm{Y}}$ readout~\cite{GBL+18}.
There, an oscillating collective magnetization induced by an RF pulse is phase-coherent,
and a free nuclear precession analogous to a free induction decay in conventional NMR has been recorded (dubbed as coherently averaged synchronized readout).
It should be noted that, for a single unpolarized nuclear spin, random nuclear orientations average out nuclear precession signals.
We detect a freely-precessing single nuclear spin by polarizing it, and in this case the signal is coherently averaged to reveal its oscillation phase.

Experiments were performed using a type-IIa (001) diamond substrate from Element Six.
Single negatively-charged NV centers in bulk ($\sim$50~$\mu$m deep) are optically resolved
by a home-built confocal microscope equipped with a 515-nm excitation laser for the initialization and readout of the NV spin.
The fluorescence from a single NV center is collected by a single-photon counting module, and the microwave to manipulate the NV spin is delivered through a copper wire running across the diamond surface.
The RF signal to control $^{13}$C nuclei (1.1\% abundance and the gyromagnetic ratio $\gamma_{\mathrm{c}}$ of 10.705~kHz/mT) is generated by a hand-wound coil bonded on the back side of the sample mount.
This configuration makes the magnetic field from the coil roughly point normal to the sample surface, but the deviation from it is carefully calibrated~\cite{SM}.
We select an NV center with its symmetry axis along the $[\bar{1}\bar{1}1]$ crystallographic direction, and $B_0$ of 36.2~mT is applied along the same direction.

\begin{figure*}
\begin{center}
\includegraphics{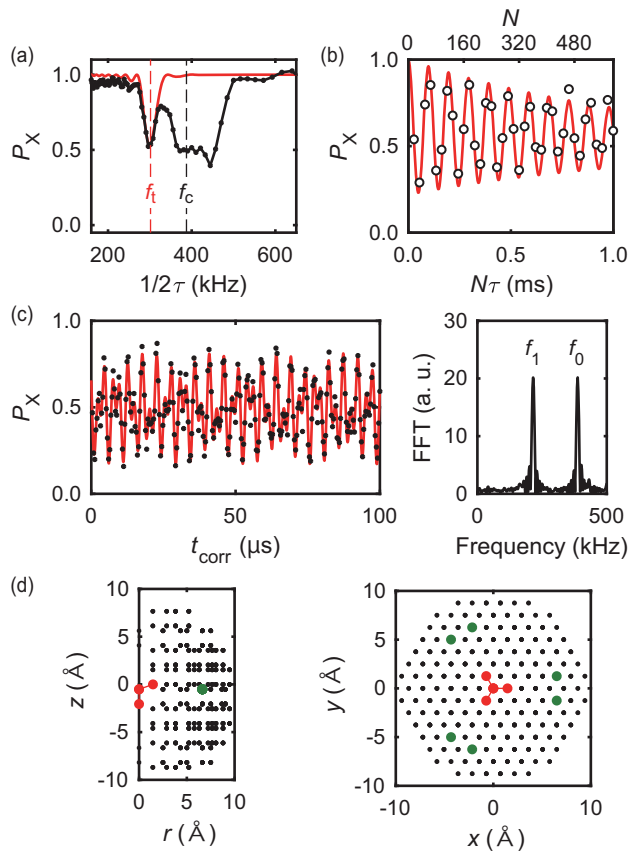}
\caption{\label{fig2}
(a) NMR spectrum taken by the CP sequence ($N$ = 16).
$f_{\mathrm{c}}$ = 387.5~kHz and $f_{\mathrm{t}}$ = 301.6~kHz.
(b) CP sequence as $N$ is incremented ($\tau$ = 1.6875~$\mu$s).
(c) Correlation spectroscopy and its Fourier transform. $f_0$ = 387.5~kHz and $f_1$ = 215.6~kHz.
In (a--c), the red solid lines are simulations using the estimated values of $A_{\parallel}$ and $A_{\perp}$.
(d) Map of the diamond lattice.
The $x$, $y$, and $z$ axes are parallel to the $[\bar{1}\bar{1}\bar{2}]$, $[1\bar{1}0]$, and $[\bar{1}\bar{1}1]$ crystallographic directions, respectively.
The black dots represent carbon sites, and the green dots are the candidate sites of the $^{13}$C nucleus examined.
It is assumed that the vacancy site locates above the nitrogen site,
and the origin is taken 0.75~\AA\, above the vacancy site~\cite{SM}.
}
\end{center}
\end{figure*}

We first characterize the magnetic environment of this NV sensor.
The primary purpose here is to find a single $^{13}$C nucleus whose coordinate parameters $r$ and $\theta$ are known
so that our protocol may be applied to determine $\phi$ and thus its exact position (lattice site).
We apply the CP sequence, in which $\tau$ is incremented with $N$ fixed as 16, and obtain the spectrum shown in Fig.~\ref{fig2}(a).
The broad dip observed around the bare $^{13}$C Larmor frequency $f_{\mathrm{c}} = \gamma_{\mathrm{c}} B_0$ = 387.5~kHz originates from weakly-coupled bath $^{13}$C nuclei,
whereas the sharp dip at $f_{\mathrm{t}}$ = 301.6~kHz is indicative of a single $^{13}$C nucleus strongly-coupled to the sensor.
The latter signal is examined in more detail by incrementing $N$ of the CP sequence with $\tau$ fixed at a near-resonance condition of 1.6875~$\mu$s = 1/(2$\times$296~kHz) [Fig.~\ref{fig2}(b)].
The $^{13}$C nuclear spin then nutates approximately about the $A_{\perp}$ axis.
For sufficiently large $N$, it makes coherent, full $2\pi$-rotations multiple times at the frequency $f_{\mathrm{cp}}$ of 10.2~kHz.
Simulations also confirm that we are observing a single nuclear spin and not multiple nuclear spins with the same hyperfine parameters~\cite{SM}.

We further analyze this $^{13}$C nucleus by correlation spectroscopy [Fig.~\ref{fig1}(c)].
Figure~\ref{fig2}(c) shows a modulated oscillation as increasing $t_{\mathrm{corr}}$,
and its Fourier transformation reveals two peaks at $f_0$ = 387.5~kHz and $f_1$ = 215.6~kHz.
$f_0$ corresponds exactly to $f_{\mathrm{c}}$, and therefore results from the coupling with $|m_S = 0 \rangle$.
On the other hand, $|m_S = -1 \rangle$ exerts the hyperfine field on the nuclear spin, shifting its precession frequency.
The target nuclear spin is properly probed because $(f_0 + f_1)/2$ = 301.55~kHz = $f_{\mathrm{t}}$, and the negative shift ($f_1 - f_0 < 0$) suggests negative $A_{\parallel}$.
Combining these results, we deduce $A_{\parallel}$ = $-$173.1~kHz and $A_{\perp}$ = 22.3~kHz with the accuracy of about 0.1~kHz~\cite{SM}.

It should be noted that, unlike the case of external nuclear spins,
$A_{\parallel}$ and $A_{\perp}$ estimated here are not purely dipolar in nature due to the presence of a relatively-strong contact hyperfine interaction.
This usually complicates the direct estimation of the coordinate parameters $r$ and $\theta$ from $A_{\parallel}$ and $A_{\perp}$.
However, the estimated values are accurate enough to be compared with theoretical calculations.
Nizovtsev {\it et al.} have recently performed an extensive density functional theory (DFT) simulation of a C$_{510}$[NV]H$_{252}$ cluster~\cite{NKP+18}.
From their list of hyperfine parameters for 510 carbon sites, we find that six sites labeled as C218, C226, C230, C240, C280, and C282 show close agreement with the experimental values.
The hyperfine parameters of these sites are, in kHz unit, ($A_{\parallel}$, $A_{\perp}$) = ($-$175.4, 21.7), ($-$176.7, 21.7), ($-$174.7, 21.7), ($-$177.1, 21.9), ($-$173, 22), and ($-$173.4, 22.1), respectively
[average: ($-$175.1$\pm$2.1, 21.9$\pm$0.2)].
The positions of these candidate sites are given in Fig.~\ref{fig2}(d).
In all cases, we obtain $r$ = 6.84~\AA~and $\theta$ = 94.8$^{\circ}$~\cite{SM}.

The next task is to polarize the target nuclear spin.
For this, we use a pulsed technique called PulsePol, which is yet another application of Hamiltonian engineering~\cite{SST+17}.
Figure~\ref{fig1}(d) shows two PulsePol sequences labeled as PolY and PolX.
One cycle consists of eight pulses with the total duration of $2\tau_{\mathrm{pol}}$.
When $2 \tau_{\mathrm{pol}} = k/f_{\mathrm{n}}$ is satisfied ($k$: odd integer, $f_{\mathrm{n}}$: nuclear Larmor frequency),
the average Hamiltonian of a hyperfine-coupled electron--nuclear system becomes proportional to
$S_{+}I_{-} + S_{-}I_{+}$ (flip--flop) or $S_{+}I_{+} + S_{-}I_{-}$ (flip--flip), driving the polarization transfer between the electron and the nuclei.
It is shown that for $k$ = 3 PolY (PolX) drives flip--flip (flip--flop), whereas for $k$ = 5 PolY (PolX) does flip--flop (flip--flip)~\cite{SM}.
Because the NV spin is optically initialized into $m_S$ = 0, PolY (PolX) for $k$ = 3 polarizes the nuclei into $m_I$ = $-\frac{1}{2}$ ($\frac{1}{2}$).
For $k$ = 5, the direction of the nuclear polarization becomes opposite. 

\begin{figure*}
\begin{center}
\includegraphics{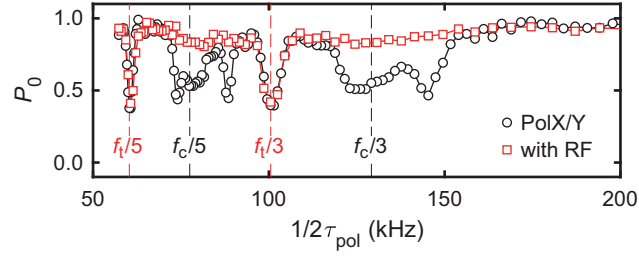}
\caption{\label{fig3}
Demonstration of PulsePol on bath nuclear spins ($\circ$) and on a single, selectively-addressed nuclear spin ($\square$).
}
\end{center}
\end{figure*}

We test the performance of PulsePol by successive application of PolX and PolY, as shown in Fig.~\ref{fig1}(e).
The first PolX serves to depolarize the nuclear polarization during a previous run,
and the polarization transfer by PolY is read out as a decrease of the $m_S$ = 0 polarization ($P_0$)~\cite{SSM+17}.
The circle ($\circ$) points in Fig.~\ref{fig3} are the result of this measurement with $N_{\mathrm{pol}}$ = 5.
Two ``replicas'' of the NMR spectrum in Fig.~\ref{fig2}(a) are clearly observed at $\frac{1}{3}$ and $\frac{1}{5}$ of the NMR conditions, as expected.
Furthermore, a single nuclear spin can be selectively polarized by applying the sequence of Fig.~\ref{fig1}(f).
This sequence works as follows.
(1) PolY is executed nine times in order to fully polarize the nuclei.
(2) A microwave $\pi$ pulse drives the NV spin into $|m_S = -1 \rangle$.
(3) An RF $\pi$ pulse tuned at $f_1$ = 215.6~kHz is applied.
The NV spin being $|m_S = -1 \rangle$, only the target nuclear spin is resonantly flipped by this RF pulse.
(4) By the final PolY, the polarization transfer acts only on the flipped nuclear spin, as other nuclei have already been polarized.
The square ($\square$) points in Fig.~\ref{fig3} clearly demonstrate the power of selective polarization;
dips are observed only at $f_{\mathrm{t}}/3$ and $f_{\mathrm{t}}/5$, with all other dips almost completely disappeared.

\begin{figure*}
\begin{center}
\includegraphics{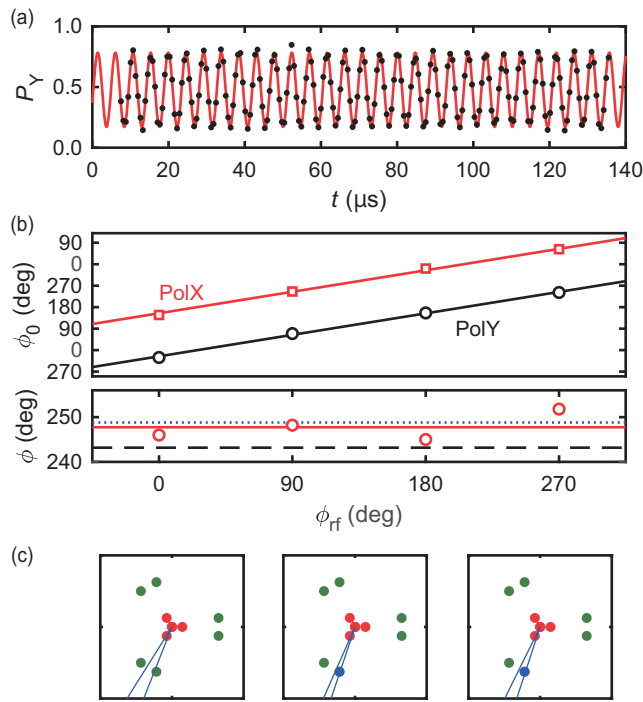}
\caption{\label{fig4}
(a) $P_{\mathrm{Y}}$ as a function of $t$ reveals a free precession of a single nuclear spin.
(b) $\phi_0$ as a function of $\phi_{\mathrm{rf}}$ (upper panel).
The solid lines are linear fits.
Estimated $\phi$ (lower panel).
(c) The blue lines indicate the accuracy ranges of $\phi$, based on three estimations of $\phi_{\mathrm{n}}(0)$.
At most one lattice site (in blue) with $\phi$ = 250.9$^{\circ}$ falls on the estimated ranges.
}
\end{center}
\end{figure*}

We now demonstrate the protocol of Fig.~\ref{fig1}(g) to determine $\phi$.
Either PolY or PolX is applied with $N_{\mathrm{pol}}$ = 5, $N_{\mathrm{rep}}$ = 5, and $\tau_{\mathrm{pol}}$ = 4.9760~$\mu$s.
$\tau_{\mathrm{pol}}$ satisfies $3/(2\tau_{\mathrm{pol}}) = f_t$, for which PolY (PolX) polarizes the target nuclear spin into $m_I$ = $-\frac{1}{2}$ ($\frac{1}{2}$).
A phase-controlled selective RF $\pi$/2 pulse tuned at $f_1$ is applied, followed by the CP sequence with the $(\pi/2)_{\mathrm{Y}}$ readout ($N$ = 16, $\tau$ = 1.6608~$\mu$s).
The RF pulse length is chosen to be 102.041~$\mu$s, matched with 22 oscillation periods (22/$f_1$),
in order to suppress unwanted phase acquisitions by the RF field along the $z$ axis~\cite{SM}.
Figure~\ref{fig4}(a) shows an exemplary trace as changing $t$.
Here, PolY is used, and the waveform of the RF signal is a cosine wave, for which we define the RF phase $\phi_{\mathrm{rf}}$ = 0$^{\circ}$.
The data is fitted by $A \cos ( 2 \pi f_{\mathrm{p}} t + \phi_0 ) + B$ (red curve).
$f_{\mathrm{p}}$ agrees well with $f_1$, confirming that the free precession of the target nuclear spin is indeed detected.

We note that the minimum $t$ is set as $t_0$ = 6.872~$\mu$s in order to avoid an overlap with an RF pulse, whereas our aim here is to estimate the oscillation phase at $t$ = 0.
To accurately estimate $\phi_0$ under this constraint, oscillations should be taken as long as possible.
This requires a dauntingly long measurement time, and at the same time exceedingly high stability of the experimental setup.
We therefore choose to undersample the data points for further measurements;
by taking less points, $t$ is instead increased up to 1~ms ($< T_1 \approx$ 5~ms),
and yet the original $\phi_0$ is recovered by appropriately setting the measurement parameters~\cite{SM}.
Figure~\ref{fig4}(b) shows $\phi_0$ determined in this way.
The linear dependence on $\phi_{\mathrm{rf}}$ is observed, as expected~\cite{SM}.
$\phi_0$ is 180$^{\circ}$-shifted between PolY and PolX, confirming that the two sequences polarize the nuclear spin into opposite directions.
From the fit, we obtain $\phi_0 = \phi_{\mathrm{rf}} + 334.0^{\circ} (\bmod \, 360^{\circ})$.
On the other hand, if we take into account the azimuthal angle of the RF field in this coordinate and the effect of detuning $(f_\mathrm{p} - f_1)$,
we can estimate $\phi_{\mathrm{n}}(0)$ as $-\phi_{\mathrm{rf}} + 89.2^{\circ}$~\cite{SM}.
Together, we obtain $\phi$ as 243.2$\pm$5.3$^{\circ}$ [dashed line in Fig.~\ref{fig4}(b)].
$\phi_{\mathrm{n}}(0)$ can be estimated more accurately by simulating the dynamics of the nuclear spin~\cite{SM}.
The simulations give $\phi$ = 248.8$\pm$2.7$^{\circ}$ (dotted line) or 247.8$\pm$4.1$^{\circ}$ (solid line), marginally dependent on the parameters used.
Figure~\ref{fig4}(c) shows the accuracy ranges of $\phi$ determined by these estimations, 
and there is at most one lattice site that falls on this range;
we have been able to pinpoint the lattice site of the target nuclear spin~\cite{SM}.

To summarize, we have described a protocol which tracks a free precession of a single nuclear spin.
Combined with DNP and a phase-controlled RF pulse, our method is capable of determining the azimuthal angle of the target nuclear spin.
A particular experimental demonstration was performed on a single $^{13}$C nuclear spin and its lattice position was uniquely pinpointed.
Previously, the position of a single $^{13}$C nuclear spin in diamond had been estimated by analyzing NMR spectra taken at three differently-oriented $\bm{B}_0$~\cite{ZHS+12}.
When the NV center or other solid-state defects with $S > \frac{1}{2}$ electron spin is used as a sensor,
an application of $\bm{B}_0$ misaligned from the sensor quantization axis complicates the analysis.
The present protocol circumvents this issue.
Looking ahead, we imagine that the present protocol could be employed for three-dimensional mapping of nuclear spins in a single molecule positioned on a near-surface NV sensor~\cite{WJGM13}.
The nuclear--nuclear interactions within the immobilized molecule can be suppressed using dipolar decoupling sequences such as WAHUHA and MREV~\cite{APN+17,H76},
which are compatible with our protocol.
The protocol can also be combined with the high-resolution spectroscopy method~\cite{GBL+18,SGS+17,BCZD17},
so that chemical shifts and J-couplings could be resolved.
Thus, our result paves the way for magnetic resonance imaging at the single-molecular level.

The authors thank C. L. Degen, J. Zopes, and J. Boss for helpful discussions, especially on preparation and calibration of an RF coil,
and also for letting us know their related work~\cite{ZCS+18}.
K.S. is supported by JSPS Grant-in-Aid for Research Fellowship for Young Scientists (DC1) No.~JP17J05890.
K.M.I is supported by JSPS Grant-in-Aid for Scientific Research (KAKENHI) (S) No.~26220602,
JST Development of Systems and Technologies for Advanced Measurement and Analysis (SENTAN),
JSPS Core-to-Core Program, and Spintronics Research Network of Japan (Spin-RNJ).

\begin{center}
{\bf Supplemental Material}
\end{center}
\tableofcontents

\section{\uppercase{D}\lowercase{efinitions of coordinates}\label{sec_1}}
\begin{figure*}
\begin{center}
\includegraphics{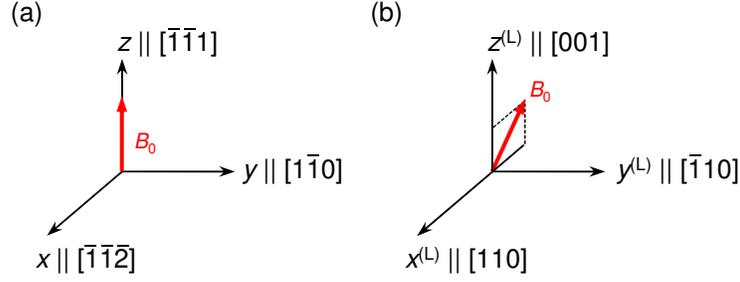}
\caption{
(a) Sensor coordinate.
(b) Laboratory coordinate.
\label{fig_s1}}
\end{center}
\end{figure*}

In the main text, the Cartesian coordinate depicted in Fig.~\ref{fig_s1}(a) was adopted.
The $x$, $y$, and $z$ axes of this coordinate are parallel to the [$\bar{1}\bar{1}\bar{2}$], [1$\bar{1}$0], and [$\bar{1}\bar{1}$1] crystallographic directions of diamond, respectively.
Since a single nitrogen-vacancy (NV) center with its symmetry axis parallel to the [$\bar{1}\bar{1}$1] direction was used as a quantum sensor
and the static external magnetic field $\bm{B}_0$ was applied along the same direction,
this coordinate is suitable to discuss the position of the target nucleus with respect to that of the sensor, and is henceforth termed as ``sensor coordinate''.
The spherical coordinate parameters ($r$, $\theta$, $\phi$)
with the distance $r$ ($\geq 0$), the polar angle $\theta$ ($0 \leq \theta \leq \pi$), and the azimuthal angle $\phi$ ($0 \leq \phi < 2\pi$) are used to specify the position of the target nucleus.

The origin is taken at an approximate ``center of mass'' of the NV center, where the NV spin is regarded as a point dipole.
It resides in the crossing point of the NV symmetry axis and the plane spanned by the three carbon atoms adjacent to the vacancy, or in other words, 0.75~\AA~above the vacancy site.
It is assumed that the vacancy site locates above the nitrogen site.
The motivation of this definition is that the direction of $A_{\perp}$ is conditional on $\theta$ and changes a sign at $\theta$ = $\frac{\pi}{2}$ [Eq.~(\ref{eq_e_perp}) in Sec.~\ref{sec_3}],
and that the target nuclear spin we measured lies coincidentally in the plane including the vacancy.
It is generally accepted that the ``center of mass'' of the NV center lies above the vacancy, even though the precise position has not been determined.
For analysis of $\phi$ of the target nuclear spin, we take the direction of $A_{\perp}$ inward.
Our definition gives $\theta$ of the target spin as 94.8$^{\circ}$, so that it is consistent with Eq.~(\ref{eq_e_perp}).

Here, we introduce ``laboratory coordinate'', the second coordinate that is to be used in this Supplemental Material.
The sensor coordinate and the laboratory coordinate are readily transformed each other,
and the purpose of using two coordinates is purely for convenience sake.
The laboratory coordinate is depicted in Fig.~\ref{fig_s1}(b).
The $x^{(\mathrm{L})}$, $y^{(\mathrm{L})}$, and $z^{(\mathrm{L})}$ axes are parallel to the [110], [$\bar{1}$10], and [001] crystallographic directions, respectively.
Because of the configuration of our experimental setup [Fig.~\ref{fig_s2}(a)],
this coordinate is suitable to discuss the direction of the magnetic field generated by a coil bonded on the back side of the sample mount (Sec.~\ref{sec_2}).
The following matrix transforms a vector $\bm{a}^{\mathrm{(L)}}$ defined in the laboratory coordinate into a vector $\bm{a}^{\mathrm{(S)}}$ in the sensor coordinate:
\begin{equation}
T^{ \mathrm{(L \rightarrow S)} } = R_y(-\Theta_0^{\mathrm{(L)}}) R_z (-\Phi_0^{\mathrm{(L)}}).
\end{equation}
$\Theta_0^{\mathrm{(L)}}$ = 54.7$^{\circ}$ and $\Phi_0^{\mathrm{(L)}}$ = 180$^{\circ}$ are the polar and azimuthal angles of $\bm{B}_0$ as seen in the laboratory coordinate, respectively.
$R_y(\Theta)$ and $R_z(\Phi)$ are the rotation matrices defined as
\begin{equation}
R_y(\Theta) = \left( 
\begin{array}{ccc}
\cos \Theta & 0 & \sin \Theta \\
0 & 1 & 0 \\
-\sin \Theta & 0 & \cos \Theta \\
\end{array} \right)
\quad\mathrm{and}\quad
R_z(\Phi) = \left( 
\begin{array}{ccc}
\cos \Phi & -\sin \Phi & 0 \\
\sin \Phi & \cos \Phi & 0 \\
0 & 0 & 1 \\
\end{array} \right).
\end{equation}

\section{\uppercase{E}\lowercase{xperimental setup}\label{sec_2}}
\subsection{Electronics}
\begin{figure*}
\begin{center}
\includegraphics{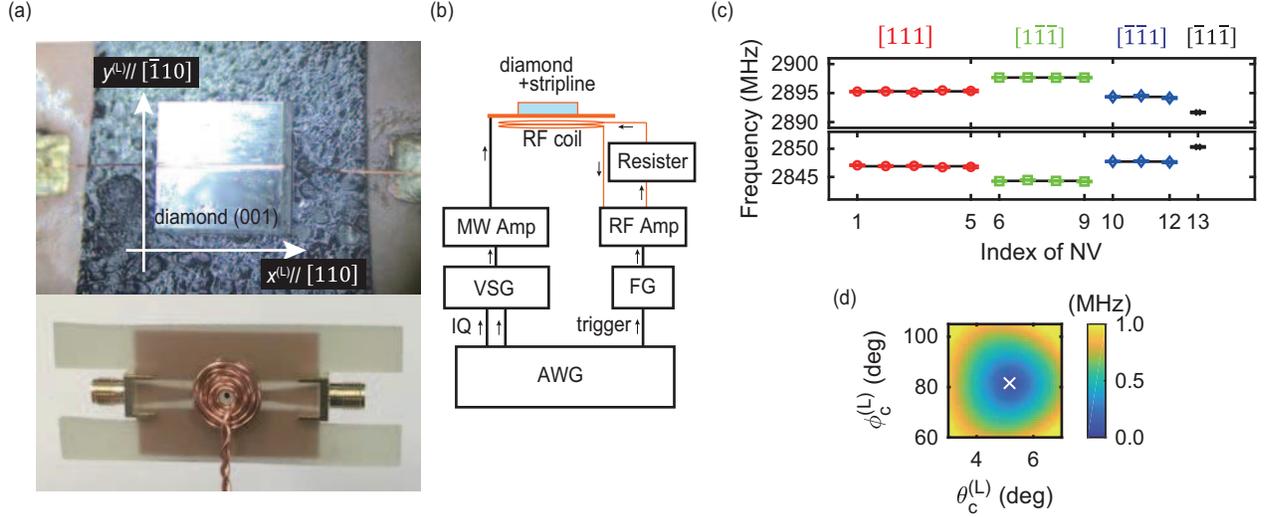}
\caption{
(a) Front and back sides of the sample mount (PCB board) with a diamond sample, a copper wire, and a coil.
(b) Schematic of electronics.
AWG: arbitrary waveform generator, VSG: vector signal generator, FG: function generator.
(c) Resonance frequencies of single NV centers found in a 80$\times$80$\times$10~$\mu$m$^3$ volume.
The black lines indicate the average resonance frequencies of the NV centers having the same symmetry axis.
(d) Determination of the direction of $\bm{B}_{\mathrm{c}}$.
The fit errors are minimized at ($\theta_{\mathrm{c}}^{\mathrm{(L)}}, \phi_{\mathrm{c}}^{\mathrm{(L)}}$) = (5.2$^{\circ}$, 81.6$^{\circ}$).
\label{fig_s2}}
\end{center}
\end{figure*}

As described in the main text, a single NV center in a type-IIa (001) diamond was measured by a home-built confocal microscope.
We here focus on the electronics aspect of our setup, which is schematically shown in Fig.~\ref{fig_s2}(b).
Microwave pulses are generated by a vector signal generator (VSG, Stanford Research Systems SG396),
amplified by a high-power broadband amplifier (Mini-Circuits ZHL-16W-43+),
and delivered to the NV center through a copper wire running across the diamond surface [Fig.~\ref{fig_s2}(a)].
An arbitrary waveform generator (AWG, Tektronix AWG7122C) provides IQ signals to the VSG.
RF pulses are triggered by the same AWG.
Waveforms generated by a function generator (FG, NF WF1973) and amplified by a low-impedance amplifier (Accel Instruments TS200-HF)
are sent to a hand-wound coil bonded on the back side of the sample mount [Fig.~\ref{fig_s2}(a)].
This coil has an inductance $L_{\mathrm{c}}$ of 3.5~$\mu$H and a resistance $R_{\mathrm{c}}$ of 0.3~$\Omega$.
A low-impedance--high-power resister ($R$ = 4.7~$\Omega$) is series-connected to the coil.
By this resistor, the rising time is reduced down to $L_{\mathrm{c}}/(R + R_{\mathrm{c}})$ = 0.7~$\mu$s, allowing us to generate RF signals at several hundreds of kHz.

\subsection{Magnetic fields generated by the coil}
To calibrate the direction of the magnetic fields generated by the coil, we conduct vector DC magnetometry using multiple single NV centers.
We optically resolve single NV centers (including the one used in the main text) in a $80\times80\times10~\mathrm{\mu m^3}$ scan volume within diamond,
and apply DC voltages on the coil to generate the DC magnetic field $\bm{B}_{\mathrm{c}}$.
Note that this $\bm{B}_{\mathrm{c}}$ is different from $\bm{B}_0$.
The latter was supplied using a permanent magnet, and in the present measurement, the magnet was removed ($B_0$ = 0~mT).
The spin resonance frequencies of 13 single NV centers under $\bm{B}_{\mathrm{c}}$ are plotted in Fig.~\ref{fig_s2}(c).
Among the NV centers having the same symmetry axis, the deviation from the averaged resonance frequency [solid lines in Fig.~\ref{fig_s2}(c)] is found to be less than 0.1~kHz,
certifying high homogeneity of $\bm{B}_{\mathrm{c}}$ within the observed volume.
The resonance frequencies of the $i$th NV center are calculated from the spin Hamiltonian
\begin{equation}
H^{(i)} = D (S_z^{(i)})^2 + \gamma_{\mathrm{e}} \bm{B}_{\mathrm{c}} \cdot \bm{S}^{(i)},
\end{equation}
where $D$ is the zero-field splitting, $\gamma_{\mathrm{e}}$ = 28~MHz/mT is the gyromagnetic ratio of the electron,
and $\bm{S}^{(i)}$ is the $S$ = 1 spin operator with its quantization axis taken as the symmetry axis of the $i$th NV center.
The direction and strength of $\bm{B}_{\mathrm{c}}$ are determined by minimizing the errors between the observed and calculated resonance frequencies [Fig.~\ref{fig_s2}(d)].
As a result, we obtain $D$ = 2870.4~MHz, $B_{\mathrm{c}}$ = 1.47~mT, $\theta_{\mathrm{c}}^{\mathrm{(L)}}$ = 5.2$^{\circ}$, and $\phi_{\mathrm{c}}^{\mathrm{(L)}}$ = 81.6$^{\circ}$,
where $\theta_{\mathrm{c}}^{\mathrm{(L)}}$ and $\phi_{\mathrm{c}}^{\mathrm{(L)}}$ are the polar and azimuthal angles of $\bm{B}_{\mathrm{c}}$ as seen in the laboratory coordinate.
In the sensor coordinate, these angles are given by $\theta_{\mathrm{c}}$ = 55.7$^{\circ}$ and $\phi_{\mathrm{c}}$ = 186.2$^{\circ}$.

\subsection{Delay time in the RF electronics}
\begin{figure*}
\begin{center}
\includegraphics{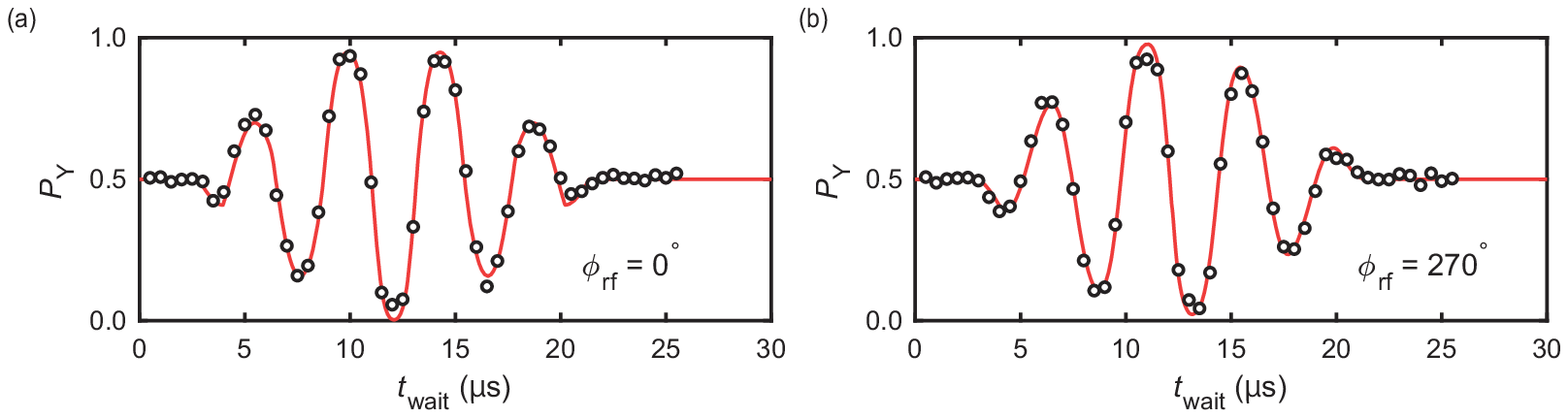}
\caption{
AC sensing for (a) $\phi_{\mathrm{rf}}$ = 0$^{\circ}$ and (b) $\phi_{\mathrm{rf}}$ = 270$^{\circ}$.
\label{fig_s3}}
\end{center}
\end{figure*}

The presence of an unknown delay directly affects the accuracy of the estimation of $\phi$.
For instance, at $f_1$ = 215.6~kHz used in our experiments, the delay of 13~ns amounts to the angle difference of $-$1$^{\circ}$.
The sources of the delay include trigger jitters in the AWG and the FG, the electrical length of the coaxial cables, the time constant of the LC circuit, and so on.
Some of them can be characterized independently, but the total delay {\it at the position of the NV center}, $t_{\mathrm{delay}}$ can only be measured using the NV center itself.
For this purpose, we use a waveform 
\begin{equation}
W(t) = \left\{
\begin{array}{ll}
V_{\mathrm{pp}} \cos [2 \pi f_1 (t - \tau_0) + \phi_{\mathrm{rf}}] & (\tau_0 \leq t \leq \tau_0 + 4 \tau) \\
& \\
0 & (\mathrm{otherwise})
\end{array} \right.
\end{equation}
with $V_{\mathrm{pp}}$ = 10~mV, $\tau_0$ = 11~$\mu$s, $\tau$ = 2.319~$\mu$s = (2$f_1$)$^{-1}$, and $\phi_{\mathrm{rf}}$ = 0$^{\circ}$ or 270$^{\circ}$.
The coil receives an amplified waveform $t_{\mathrm{delay}}$ seconds after the trigger of the AWG.
At the same time, the CP sequence with $N$ = 4 and $\tau$ = 2.319~$\mu$s, starting $t_{\mathrm{wait}}$ seconds after the trigger, detects this AC field.
When read out by the ($\pi$/2)$_{\mathrm{Y}}$ pulse, the transition probability is given by~\cite{SGS+17s}
\begin{equation}
P_{\mathrm{Y}} = \frac{1}{2}(1-\sin \varphi)
\label{py_rf}
\end{equation}
with
\begin{equation}
\varphi = \frac{ 2 \pi \gamma_{\mathrm{e}} b_{\mathrm{rf}} }{ V_{\mathrm{pp}} } \int W(t - t_{\mathrm{delay}}) y(t-t_{\mathrm{wait}}) dt.
\label{eq_varphi}
\end{equation}
$b_{\mathrm{rf}}$ is the AC signal amplitude at the position of the NV center, and $y(t)$ is the modulation function
\begin{equation}
y(t) = \left\{
\begin{array}{ll}
1 & ( 0 \leq t < \frac{\tau}{2}, \frac{3 \tau}{2} \leq t < \frac{5 \tau}{2}, \frac{7 \tau}{2} \leq t < 4 \tau ) \\
& \\
-1 & ( \frac{\tau}{2} \leq t < \frac{3 \tau}{2}, \frac{5 \tau}{2} \leq t < \frac{7 \tau}{2} ) \\
& \\
0 & (\mathrm{otherwise}).
\end{array} \right.
\end{equation}
Equation~(\ref{eq_varphi}) means that the accumulated phase $\varphi$ is a convolution of a signal wave with delay $t_{\mathrm{delay}}$ and a ``sensing window'' of the CP sequence.
When the timing of the CP sequence matches with that of the signal wave, $P_{\mathrm{Y}}$ is modified accordingly.
Therefore, by sweeping $t_{\mathrm{wait}}$, we can estimate $t_{\mathrm{delay}}$.
Figure~\ref{fig_s3} shows $P_{\mathrm{Y}}$ as a function of $t_{\mathrm{wait}}$ for $\phi_{\mathrm{rf}}$ = 0$^{\circ}$ (a) and 270$^{\circ}$ (b).
The data is fitted by Eq.~(\ref{py_rf}), and we extract $t_{\mathrm{delay}}$ = 1.088$\pm$0.003~$\mu$s.
The error of $\pm$3~ns corresponds to only $\pm$0.2$^{\circ}$.

\subsection{Chirped microwave pulse}
In our experiments based on the pulse sequences shown in Fig.~1 of the main text, single photons emitted from the NV center are counted by a single-photon counting module (Laser Components COUNT-10C).
These events are indicated as {\it Readout} in Fig.~1 of the main text.
The recorded photon counts are converted into the ``transition probabilities'' (Sec.~\ref{sec_3}).
For accurate conversion, calibration of the photon counts marking $|m_S = 0 \rangle$ and $|m_S = -1 \rangle$ is crucial. 
To robustly flip the NV spin from laser-initialized $|m_S = 0 \rangle$ to $|m_S = -1 \rangle$,
we use a chirped microwave pulse known as WURST (wideband, uniform rate, smooth truncation)~\cite{KF95s,SSB+16s}.
The amplitude modulation of WURST is given by
\begin{eqnarray}
\left( 1 - \left| \cos \frac{\pi t}{ t_{\mathrm{p}} } \right| \right)^{ \alpha_{\mathrm{p}} },
\end{eqnarray}
where $t_{\mathrm{p}}$ is the pulse length and $\alpha_{\mathrm{p}}$ characterizes the envelope shape.
We set $t_{\mathrm{p}}$ = 2~$\mu$s and $\alpha_{\mathrm{p}}$ = 2, and sweep the microwave frequency from $-$10 to 10~MHz around the resonance frequency during the pulse.
The pulse shaping and frequency modulation are done by the AWG.
Prior to the respective pulse sequences, we record the reference photon counts with and without a chirped pulse.

A chirped pulse is also used in the protocol of Fig.~1(g) of the main text, when the NV spin is prepared in $|m_S = -1 \rangle$ before applying an RF $\pi$/2 pulse (Sec.~\ref{sec_8}).
Note that microwave pulses in the CP and PulsePol sequences are {\it not} chirped.

\section{\uppercase{D}\lowercase{ynamics of a single nuclear spin under the }\uppercase{CP}\lowercase{ sequence}\label{sec_3}}
Here, we derive Eqs.~(1) and (2) of the main text, closely following the descriptions in~\cite{TWS+12s} and~\cite{BCA+16s}.
In the sensor coordinate, the spin Hamiltonian of the NV--$^{13}$C-coupled system is given by
\begin{equation}
H = D S_z^2 + \gamma_{\mathrm{e}} B_0 S_z - \gamma_{\mathrm{c}} B_0 I_z + S_z (A_{\perp} \bm{e}_{\perp} \cdot \bm{I} + A_{\parallel} I_z).
\label{eq_h}
\end{equation}
The target $^{13}$C nuclear spin is located at ($r$, $\theta$, $\phi$).
$\bm{e}_{\perp}$ is given by
\begin{equation}
\bm{e}_{\perp} = \left\{
\begin{array}{ll}
\cos \phi \, \bm{e}_x + \sin \phi \, \bm{e}_y & ( 0 \leq \theta < \frac{\pi}{2} ) \\
& \\
-\cos \phi \, \bm{e}_x - \sin \phi \, \bm{e}_y & ( \frac{\pi}{2} \leq \theta \leq \pi ).
\end{array} \right.
\label{eq_e_perp}
\end{equation}
The direction of $A_{\perp}$ is conditional on $\theta$; when the nuclear spin locates above (below) the sensor spin, $A_{\perp}$ points outward (inward)
because $A_{\perp} \propto 3 \cos \theta \sin \theta/r^3$ in a dipolarly-coupled system.
Figure~1(a) of the main text depicts the case for $0 \leq \theta \leq \frac{\pi}{2}$.
The operator $\bm{e}_{\perp} \cdot \bm{I}$ can be simplified as $I_x$ by an appropriate unitary transformation (a rotation about the $z$ axis),
so that the eigenenergies of Eq.~(\ref{eq_h}) are independent of $\phi$.
However, we keep this form in order to examine how the real-space position of the $^{13}$C nuclear spin is reflected in its dynamics.

In our experiments, $|m_S = 0 \rangle$ and $|m_S = -1 \rangle$ of the NV spin, separated by $D - \gamma_{\mathrm{e}} B_0$ = 1.8582~GHz,
are used as $|0 \rangle$ and $|1 \rangle$ of the sensor, respectively.
(Strictly speaking, $^{14}$N isotope of the NV center has a nuclear spin $I$ = 1, and the $m_I$ = 1 state we used has the transition frequency 2~MHz higher than $D - \gamma_{\mathrm{e}} B_0$,
due to the NV--$^{14}$N hyperfine interaction.
In addition, the $^{14}$N nuclear spin is polarized into the $m_I$ = 1 state by optical hyperpolarization~\cite{AS18s}.
Other nuclear sublevels thus do not play any roles in the present work, and are not considered below.)
We move to the rotating frame of the NV spin and rewrite Eq.~(\ref{eq_h}) as
\begin{equation}
H_{\mathrm{r}} = -\gamma_{\mathrm{c}} B_0 I_z + S_z (A_{\perp} \bm{e}_{\perp} \cdot \bm{I} + A_{\parallel} I_z).
\label{eq_hr}
\end{equation}
Note that the nuclear spin stays in the original sensor coordinate.
When the NV spin is $|m_S = 0 \rangle$, $H_{\mathrm{r}}$ is reduced to
\begin{equation}
H_0 = -\gamma_{\mathrm{c}} B_0 I_z = -f_0 I_z.
\end{equation}
For $|m_S = -1\rangle$, we obtain
\begin{equation}
H_1 = -f_0 I_z - (A_{\perp} \bm{e}_{\perp} \cdot \bm{I} + A_{\parallel} I_z) = -f_1\bm{e}_{\mathrm{p}} \cdot \bm{I},
\end{equation}
with
\begin{eqnarray}
f_1 &=& \sqrt{(f_0 + A_{\parallel})^2 + A_{\perp}^2} \label{eq_f1} \\
\bm{e}_{\mathrm{p}} &=& \frac{ f_0 + A_{\parallel} }{f_1} \bm{e}_z + \frac{ A_{\perp} }{f_1} \bm{e}_{\perp} = \cos \theta_{\mathrm{p}} \bm{e}_z + \sin \theta_{\mathrm{p}} \bm{e}_{\perp}. \label{eq_e_p}
\end{eqnarray}
In our experiments, the polar angle is calculated as $\theta_{\mathrm{p}} = \arctan[A_{\perp}/(f_0 + A_{\parallel})]$ = 5.9$^{\circ}$.
In the protocol of Fig.~1(g) of the main text, the nuclear spin is driven by an RF field while the sensor is $|m_S = -1\rangle$.
Therefore, $\bm{e}_{\mathrm{p}}$ is the precession axis of the target nuclear spin (hence the suffix ``p'').
We detail this point in Sec.~\ref{sec_8}.

We now examine the dynamics of a single nuclear spin under the CP sequence by calculating the transition probabilities, which are the probabilities that, at the end of the CP sequence,
the NV spin is found to be in the state opposite to the state right before the application of the first $(\pi/2)_{\mathrm{X}}$ pulse.
In our experiments, both $|m_S = 0 \rangle$ and $|m_S = -1 \rangle$ are used as the initial state of the CP sequence,
but the expressions of the transition probabilities obtained below [Eqs.~(\ref{eq_px}) and (\ref{eq_py})] do not depend on the initial state.
We assume that the sensor is initialized to $|m_S = 0 \rangle$ with its pure state density matrix given by $|0 \rangle \langle 0|$.
We also introduce a density matrix of the nuclear spin as
\begin{equation}
\rho_{\mathrm{n}} = \frac{1}{2} \bm{1} + \bm{\nu}_{\mathrm{n}} \cdot \bm{I},
\end{equation}
where $\bm{\nu}_{\mathrm{n}}$ is the Bloch vector of the nuclear spin defined in the sensor coordinate.
Again, we assume that the nuclear spin is in the pure state, which is relevant because the nuclear spin is polarized in our protocol, and the coherence time of the nuclear spin is long.
Therefore, $\bm{\nu}_{\mathrm{n}}$ is parametrized by $(\theta_{\mathrm{n}}, \phi_{\mathrm{n}})$.
Note that both $\theta_{\mathrm{n}}$ and $\phi_{\mathrm{n}}$ are time-dependent in the sensor coordinate.
The density matrix of an uncoupled NV--$^{13}$C system is written as 
\begin{equation}
\rho_0 = |0 \rangle \langle 0| \otimes \rho_{\mathrm{n}}.
\end{equation}

The $(\pi/2)_{\mathrm{X}}$ pulse is applied first to create a superposition of $|0 \rangle$ and $|1 \rangle$, which couple with the nuclear spin differently.
The unitary operator for the $(\pi/2)_{\mathrm{X}}$ pulse is given as
\begin{equation}
U_{\mathrm{X}} = \frac{1}{\sqrt{2}} \left( |0 \rangle \langle 0| + |1 \rangle \langle 1| - i |1 \rangle \langle 0| -i |0 \rangle \langle 1| \right).
\end{equation}
The evolution of the nuclear spin during the CP sequence is described as
\begin{equation}
U_{\mathrm{cp}} = (U_0 U_1^2 U_0)^{ \frac{N}{2} } |0 \rangle \langle 0| + (U_1 U_0^2 U_1)^{ \frac{N}{2} } |1 \rangle \langle 1|
\label{eq_u_cp}
\end{equation}
with
\begin{eqnarray}
U_0 &=& e^{ -2\pi i H_0 \frac{\tau}{2} } = e^{ i \pi f_0 \tau I_z} = e^{ i \alpha I_z} \label{eq_u0} \\
U_1 &=& e^{ -2\pi i H_1 \frac{\tau}{2} } = e^{ i \pi f_1 \tau \bm{e}_{\mathrm{p}} \cdot \bm{I} } = e^{ i \beta \bm{e}_{\mathrm{p}} \cdot \bm{I} }. \label{eq_u1}
\end{eqnarray}
By noting that any unitary evolution of a single spin can be described as a rotation around a certain axis, we can rewrite Eq.~(\ref{eq_u_cp}) as
\begin{equation}
U_{\mathrm{cp}} = e^{ -i N \phi_{\mathrm{cp}} \bm{n}_0 \cdot \bm{I} } |0 \rangle \langle 0| + e^{ -i N \phi_{\mathrm{cp}} \bm{n}_1 \cdot \bm{I} } |1 \rangle \langle 1|,
\label{eq_u_cp_2}
\end{equation}
where $\bm{n}_0$ and $\bm{n}_1$ define the rotation axes of the respective unitary operations, and $N \phi_{\mathrm{cp}}$ is the rotation angle.
From straightforward calculations, we obtain $\phi_{\mathrm{cp}}$ as
\begin{equation}
\cos \phi_{\mathrm{cp}} = \cos \alpha \cos \beta - \cos \theta_{\mathrm{p}} \sin \alpha \sin \beta.
\label{eq_cos_phi_cp}
\end{equation}
with $\alpha = \pi f_0 \tau$ and $\beta = \pi f_1\tau$ [as defined by Eqs.~(\ref{eq_u0}) and (\ref{eq_u1})].
For later convenience, we also give an explicit form of $\bm{n}_0$:
\begin{equation}
\bm{n}_0 =
-\frac{ \sin \theta_{\mathrm{p}} \sin \beta }{ \sin \phi_{\mathrm{cp}} } \bm{e}_{\perp}
-\frac{ \cos \theta_{\mathrm{p}} (\sin \alpha \cos \beta + \cos \theta_{\mathrm{p}} \cos \alpha \sin \beta ) }{ \sin \phi_{\mathrm{cp}} } \bm{e}_z.
\label{eq_n0}
\end{equation}

The transition probability with the $(\pi/2)_{\mathrm{X}}$ readout pulse is calculated as
\begin{equation}
P_{\mathrm{X}} = \mathrm{Tr}[ (S_z^2) U_{\mathrm{X}} U_{\mathrm{cp}} U_{\mathrm{X}} \rho_0 U_{\mathrm{X}}^{\dagger} U_{\mathrm{cp}}^{\dagger} U_{\mathrm{X}}^{\dagger} ].
= 1 - \frac{1}{2 }(1-\bm{n}_0 \cdot \bm{n}_1) \sin^2 \frac{N \phi_{\mathrm{cp}}}{2}.
\label{eq_px}
\end{equation}
This is Eq.~(1) of the main text.
When the $(\pi/2)_{\mathrm{Y}}$ pulse is used, the transition probability becomes
\begin{eqnarray}
P_{\mathrm{Y}} &=& \mathrm{Tr}[ (S_z^2) U_{\mathrm{Y}} U_{\mathrm{cp}} U_{\mathrm{X}} \rho_0 U_{\mathrm{X}}^{\dagger} U_{\mathrm{cp}}^{\dagger} U_{\mathrm{Y}}^{\dagger} ] \nonumber \\
&=& \frac{1}{2} + \frac{1}{4} \bm{\nu}_n \cdot \left\{ (\bm{n}_0 - \bm{n}_1) \sin (N \phi_{\mathrm{cp}}) +2 (\bm{n}_0 \times \bm{n}_1) \sin^2 \frac{ N \phi_{\mathrm{cp}} }{2} \right\}.
\label{eq_py}
\end{eqnarray}
$U_{\mathrm{Y}}$, the unitary operator for the $(\pi/2)_{\mathrm{Y}}$ pulse, is given as
\begin{equation}
U_{\mathrm{Y}} = \frac{1}{\sqrt{2}} \left( |0 \rangle \langle 0| + |1 \rangle \langle 1| + |1 \rangle \langle 0| - |0 \rangle \langle 1| \right).
\end{equation}

Up to this point, calculations are rigorous and general.
We now consider a situation appropriate to our experiments.
Both $P_{\mathrm{X}}$ and $P_{\mathrm{Y}}$ are strongly modulated when $\bm{n}_0$ and $\bm{n}_1$ are anti-parallel {\it i.e.,} $\bm{n}_0 \cdot \bm{n}_1 = -1$.
This also means $\bm{n}_1 = -\bm{n}_0$ and $\bm{n}_0 \times \bm{n}_1 = 0$.
Equation~(\ref{eq_py}) is simplified as
\begin{equation}
P_{\mathrm{Y}} = \frac{1}{2} + \frac{1}{2} \bm{\nu}_{\mathrm{n}} \cdot \bm{n}_0 \sin (N \phi_{\mathrm{cp}}).
\label{eq_py_2}
\end{equation}

In the high-field regime ($f_0 \gg |A_{\parallel}|, A_{\perp}$),
the condition $\bm{n}_0 \cdot \bm{n}_1 = -1$ is realized, for instance, by choosing $\tau$ such that~\cite{TWS+12s}
\begin{equation}
2\tau \approx \frac{1}{ f_0 + A_{\parallel}/2 } \approx \frac{1}{f_{\mathrm{t}}},
\end{equation}
which is also the setting of our protocol.
In this case, we evaluate $\alpha \approx \pi f_0 /(f_0 + A_{\parallel}/2) \approx \pi/2$, $\beta \approx \pi f_1 /(f_0 + A_{\parallel}/2) \approx \pi/2$, and $\cos \theta_{\mathrm{p}} \approx 1$.
From Eq.~(\ref{eq_cos_phi_cp}), we obtain $\cos \phi_{\mathrm{cp}} \approx \cos \theta_{\mathrm{p}} \approx -1$, and hence $\phi_{\mathrm{cp}} \approx \pi$.
It should be noted that we can assume either $\phi_{\mathrm{cp}} > \pi$ or $\phi_{\mathrm{cp}} < \pi$~\cite{BCA+16s}.
This arbitrariness affects the directions of $\bm{n}_{0,1}$ and a sign of $\sin (N \phi_{\mathrm{cp}})$, but does not affect the final results, because $U_{\mathrm{cp}}$ is identical in both cases.
We set $\phi_{\mathrm{cp}} \approx \pi - \theta_{\mathrm{p}} < \pi$.
In this case, for sufficiently small and even $N$, $\sin (N \phi_{\mathrm{cp}}) < 0$.
From~(\ref{eq_n0}), and by noting $(\sin \theta_{\mathrm{p}} / \sin \phi_{\mathrm{cp}}) \approx 1$, we obtain $\bm{n}_0 \approx -\bm{e}_{\perp}$.
Moreover, after the application of the RF $\pi/2$ pulse, the nuclear spin is also in the $xy$ plane ($\theta_{\mathrm{n}}$ = $\pi/2$).
Equation~(\ref{eq_py_2}) becomes
\begin{equation}
P_{\mathrm{Y}} = \left\{
\begin{array}{ll}
\frac{1}{2} - \frac{1}{2} \cos (\phi - \phi_{\mathrm{n}}) \sin (N \phi_{\mathrm{cp}}) & ( 0 \leq \theta < \frac{\pi}{2} ) \\
& \\
\frac{1}{2} - \frac{1}{2} \cos (\phi - \phi_{\mathrm{n}} + \pi) \sin (N \phi_{\mathrm{cp}}) & ( \frac{\pi}{2} \leq \theta \leq \pi ).
\end{array} \right.
\label{eq_py_3}
\end{equation}
which is Eq.~(2) of the main text.

\section{\uppercase{E}\lowercase{stimation of hyperfine parameters}\label{sec_4}}
From Eqs.~(\ref{eq_f1}) and (\ref{eq_cos_phi_cp}), we obtain
\begin{eqnarray}
A_{\parallel} &=& \frac{ \cos \alpha \cos \beta - \cos (\pi-2\pi f_{\mathrm{cp}} \tau) }{ \sin \alpha \sin \beta} f_1 - f_0 \label{eq_A_para} \\
A_{\perp} &=& \sqrt{ f_1^2 - (f_0 + A_{\parallel})^2 }. \label{eq_A_perp}
\end{eqnarray}
$f_{\mathrm{cp}}$ was determined in Fig.~2(b) of the main text.
We relate $f_{\mathrm{cp}}$ and $\phi_{\mathrm{cp}}$ as $\phi_{\mathrm{cp}} = \pi - 2\pi f_{\mathrm{cp}} \tau$.
$f_0$ and $f_1$ were determined in Fig.~2(c) of the main text.
Substituting $f_{\mathrm{cp}}$ = 10.2~kHz, $f_0$ = 387.5~kHz, and $f_1$ = 215.6~kHz, and $\tau$ = 1.6875~$\mu$s into Eqs.~(\ref{eq_A_para}) and (\ref{eq_A_perp}),
we deduce $A_{\parallel}$ = $-$173.1~kHz and $A_{\perp}$ = 22.3~kHz as given in the main text.

\section{\uppercase{T}\lowercase{he number of nuclear spins contributing to the signal}\label{sec_5}}
\begin{figure*}
\begin{center}
\includegraphics{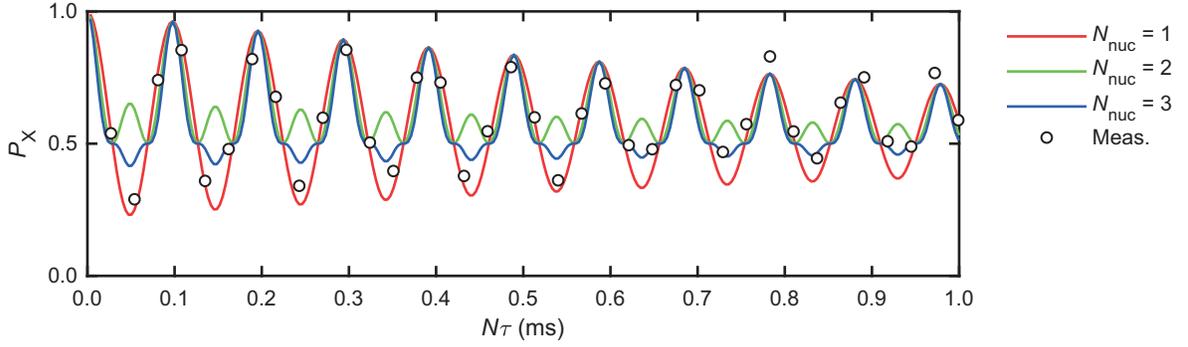}
\caption{
Data in Fig.~2(b) of the main text, together with simulations for $N_{\mathrm{nuc}}$ = 1, 2, and 3.
A single exponential decay with the time constant of 1.23~ms is superposed.
\label{fig_s4}}
\end{center}
\end{figure*}

Even though $A_{\parallel}$ and $A_{\perp}$ are determined with high precisions, there is still a possibility that multiple nuclear spins are contributing to the experimental data.
This is possible if multiple nuclear spins share the same hyperfine parameters (within experimental accuracy),
This occurs, for instance, if they occupy lattice sites equivalent to each other by symmetry. 
When multiple nuclear spins are involved but can be regarded as independent, $P_{\mathrm{X}} $ is given as 
\begin{equation}
P_{\mathrm{X}} = \frac{1}{2} [1 + \prod_{i = 1}^{N_{\mathrm{nuc}}}(2P_{\mathrm{X},i} - 1)],
\label{eq_P_X_N}
\end{equation}
where $N_{\mathrm{nuc}}$ is the number of nuclear spins.
Figure~\ref{fig_s4} shows the data in Fig.~2(b) of the main text, together with simulations ($N_{\mathrm{nuc}}$ = 1, 2, and 3) performed using the experimental values.
A large oscillation amplitude with its minimum less than 0.5 and a single-component oscillation are hallmarks of a single nuclear spin.
The experimental data clearly shows these features, especially at early times before the damping occurs. 

\section{\uppercase{P}\lowercase{ulse}\uppercase{P}\lowercase{ol method and calibration of the }\uppercase{RF}\lowercase{ pulse length}\label{sec_6}}
In the main text, PulsePol, a pulsed DNP technique recently developed by Schwartz {\it et al.}~\cite{SST+17s},
was used for the polarization transfer between the NV electron spin and the $^{13}$C nuclear spins.
While other DNP techniques such as NOVEL~\cite{HDSW88s} and optical pumping~\cite{SSC+16s} may also be applied to this system,
we find PulsePol particularly useful because of its robustness and control flexibility.
Even at low magnetic fields used in our experiments ($B_0$ = 36.2~mT), highly efficient polarization transfer was achieved.
In addition, the direction of the polarization can be controlled.
Here, we elaborate these features of PulsePol by simulations.

\begin{figure*}
\begin{center}
\includegraphics{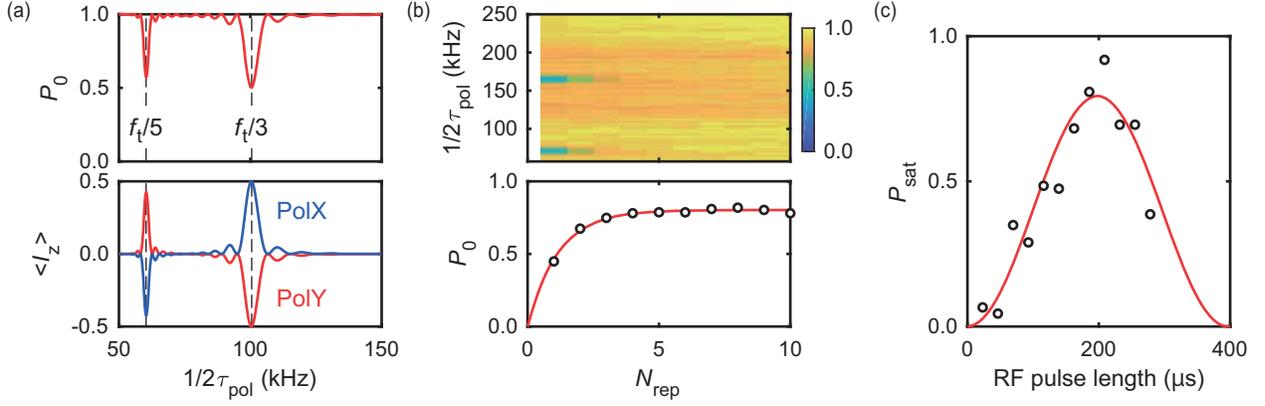}
\caption{
(a) Simulations of the polarization transfer (upper) and the nuclear spin state (lower) after PolX/Y with $N_{\mathrm{pol}}$ = 5.
(b) Selective polarization transfer by PulsePol as a function of $N_{\mathrm{rep}}$.
(c) $P_{\mathrm{sat}}$ at $k$ = 3 as a function of the RF pulse length.
\label{fig_s5}}
\end{center}
\end{figure*}

In Fig.~\ref{fig_s5}(a), we simulate the polarization transfer signal (upper) and the nuclear spin state (lower) after PolX/Y with $N_{\mathrm{pol}}$ = 5.
The initial state of the target nuclear spin is assumed to be completely mixed ($\rho_{\mathrm{n}} = \frac{1}{2} \bm{1}$).
The NV spin is initialized to $|m_S = 0 \rangle$, and the probability of staying in $|m_S = 0 \rangle$, $P_0$, is reduced when the polarization transfer occurs.
Simulated $P_0$ reproduces well with the experimental observation shown in Fig.~3 ($\square$) of the main text.
Simulated $\langle I_z \rangle$ shows that the polarization direction is controllable by $k$ (= $2 \tau_{\mathrm{pol}}/f_{\mathrm{t}}$) and the phase-cycling of the sequence (PolX or PolY).
$N_{\mathrm{pol}}$ is a tunable parameter that determines the bandwidth.
We used $N_{\mathrm{pol}}$ = 5 to selectively polarize the target nuclei at $k$ = 3 by tuning the bandwidth close to that of the $N$ =16 CP sequence.

As described in~\cite{SST+17s}, the rate of polarization transfer is maximized at $k$ = 3.
The number of repetitions required to transfer an angular momentum of 1 ($ \times \hbar$) is estimated as
$N_{\mathrm{rep}} \geq \pi f_{\mathrm{t}}/(3 (2 + \sqrt{2}) A_{\perp} N_{\mathrm{pol}}) \sim$ 2.5,
where $f_{\mathrm{t}}$ = 301.6~kHz and $A_{\perp}$ = 22.3~kHz are the experimental values.
The color plot in Fig.~\ref{fig_s5}(b) shows the evolution of the polarization transfer as $N_{\mathrm{rep}}$ is increased.
The spectrum for $N_{\mathrm{rep}}$ = 1 corresponds to the square ($\square$) points of Fig.~3 of the main text.
The transferred polarization is evaluated by integrating the $k$ = 3 dip [the lower panel of Fig.~\ref{fig_s5}(b)].
The transfer is efficient up to $N_{\mathrm{rep}}$ = 3, consistent with the estimation above.

The RF pulse length was calibrated by observing the saturated polarization $P_{\mathrm{sat}}$ as a function of the RF pulse length [Fig.~\ref{fig_s5}(c)].
In order to accurately track the motion of the nuclear spin, we want the lengths of the RF pulses to be integer multiples of oscillation periods.
We set the length of the RF $\pi$ ($\pi$/2) pulse as 199.443~$\mu$s (102.041~$\mu$s), corresponding to 43 (22) oscillation periods of $f_1$ = 215.6~kHz.

\section{\uppercase{A}\lowercase{nalysis of undersampled oscillations}\label{sec_7}}
\begin{figure*}
\begin{center}
\includegraphics{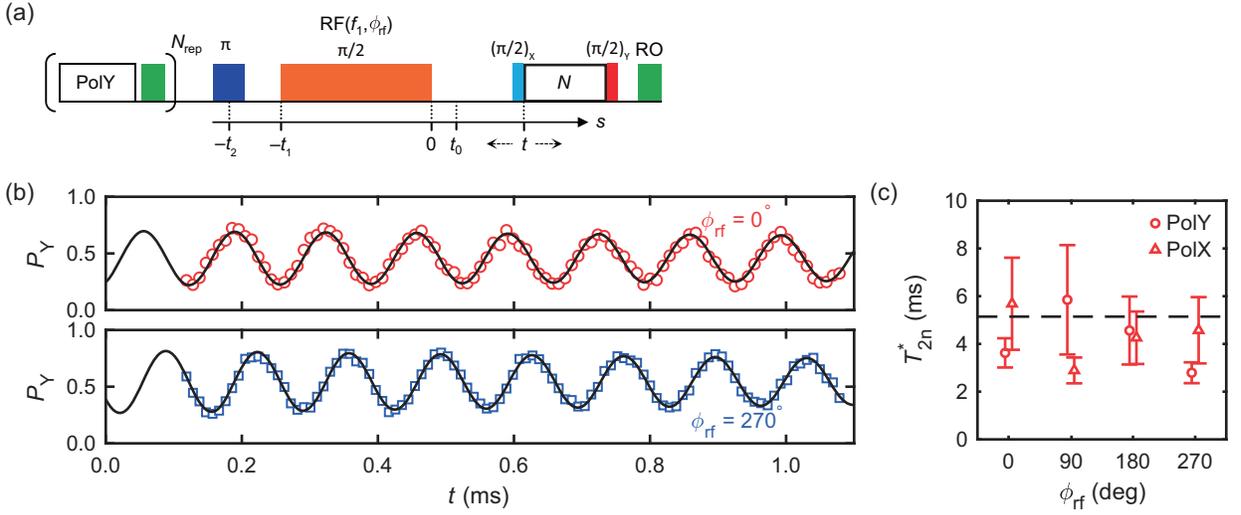}
\caption{
(a) Pulse protocol to observe nuclear free precessions.
See also Fig.~1(f) of the main text.
(b) Undersampled nuclear free precessions.
Solid lines are fits by $A e^{-t/T_{\mathrm{2n}}^*} \cos (2\pi f^{(4)} t + \alpha^{(4)}) + B$.
(c) $T_{\mathrm{2n}}^{*}$ of the measured nuclear free precessions.
$\circ$: PolY, $\triangle$: PoX.
The dashed line indicates $T_1$ of the sensor spin.
\label{fig_s6}}
\end{center}
\end{figure*}

The data in Fig.~4(b) of the main text was intentionally undersampled in order to secure sufficiently long $t >$ 1~ms [Fig.~\ref{fig_s6}(a)].
We outline a procedure to analyze undersampled data.
In signal processing, the sampling theorem asserts that an oscillation at frequency $f$ can be recovered if the sampling rate $(\Delta t)^{-1}$ is set to satisfy
\begin{equation}
\frac{1}{\Delta t} \geq 2f \,\,\,{\mathrm{or}}\,\,\, f_{\mathrm{N}} \equiv \frac{1}{2 \Delta t} \geq f,
\end{equation}
where $f_{\mathrm{N}}$ is called the Nyquist frequency.
When this condition is not met, measurement points are undersampled.
There is a non-zero integer $m$ that satisfies
\begin{equation}
m f_{\mathrm{N}} \leq f < (m + 1) f_{\mathrm{N}}.
\end{equation}
For even $m$, $f$ and the frequency obtained by undersampling, $f^{(m)}$, is connected by
\begin{eqnarray}
f^{(m)} = f - m f_{\mathrm{N}}.
\end{eqnarray}
Suppose that by fitting to the undersampled data we obtain the phase $\eta^{(m)}$ but the original oscillation has the phase $\eta$.
The following relation must be satisfied for any integer $k$:
\begin{equation}
2\pi f^{(m)} (t_0 + k \Delta t) + \eta^{(m)} = 2\pi f ( t_0 + k \Delta t ) + \eta,
\end{equation}
where $t_0$ is the starting time of the sampling.
It follows that $\eta = \eta^{(m)} - 2\pi \, m f_{\mathrm{N}} (t_0 + k \Delta t)$, and therefore
\begin{equation}
\eta = \eta^{(m)} - m \frac{\pi t_0}{\Delta t} \quad (\bmod \,\, 2 \pi).
\label{eq_eta}
\end{equation}
From Fig.~4(a) of the main text, we already know the oscillation frequency to be 216~kHz.
The frequency resolution is limited by relatively short $t$ of about 100~$\mu$s,
but is sufficient to determine an appropriate undersampling condition.
We chose $\Delta t$ = 9.600~$\mu$s for undersampling of $m$ = 4,
and $t_0$ = 6.872~$\mu$s in order to account for the RF pulse length.
We can then recover the original phase using Eq.~(\ref{eq_eta}).

Figure~\ref{fig_s6}(b) shows undersampled nuclear free precessions for $\phi_{\mathrm{rf}}$ = 0$^{\circ}$ and 270$^{\circ}$ (see Sec.~\ref{sec_8} for the definition of $\phi_{\mathrm{rf}}$).
From fits to the data, we obtain 7.5~kHz, consistent with undersampling of $m$ = 4.
The recovered original frequency is 215.8~kHz.
The phases determined by Eq.~(\ref{eq_eta}) are plotted in Fig.~4(b) of the main text.
Even though $t$ is extended longer than 1~ms, no significant decays are observed, as expected for nuclear spins.
The decay times $T_{\mathrm{2n}}^*$ are plotted in Fig.~\ref{fig_s6}(c).
They all take similar values, and fall on around $T_1$ = 5.1~ms of the sensor spin.
It is likely that true $T_{2\mathrm{n}}^*$ could be longer.
Nonetheless, $t$ can still be extended up to $T_1$ to achieve better precisions.

\section{\uppercase{D}\lowercase{etermination of $\phi$}\label{sec_8}}
In our demonstration shown in Fig.~4(a) of the main text, the origin of the time axis is defined as the end time of the RF pulse [Fig.~\ref{fig_s6}(a)],
after which the CP sequence to detect nuclear spin precessions can be applied (otherwise the RF field much stronger than the nuclear spin signal will be detected, as performed in Sec.~\ref{sec_2}).
The crucial information is the azimuthal angle of $\bm{\nu}_{\mathrm{n}}(s = 0)$, $\phi_{\mathrm{n}}(0)$.
If $\phi_{\mathrm{n}}(0)$ is known and the initial phase of the free precession is experimentally determined as $\phi_0$,
Eq.~(\ref{eq_py_3}) allows us to determine $\phi$ as
\begin{equation}
\phi = \left\{
\begin{array}{ll}
\phi_{\mathrm{n}}(0) + \phi_0 & ( 0 \leq \theta < \frac{\pi}{2} ) \\
& \\
\phi_{\mathrm{n}}(0) + \phi_0 + \pi & ( \frac{\pi}{2} \leq \theta \leq \pi ).
\end{array} \right.
\label{eq_phi}
\end{equation}
Since we have determined the polar angle $\theta$ of the target nuclear spin to be 94.8$^{\circ}$ in the main text, the second relation applies to our case.
In future applications aiming at single-molecular NMR spectroscopy, target molecules will be placed at a diamond surface, and the polar angles of target nuclei will always be less than $\pi/2$.

\begin{figure*}
\begin{center}
\includegraphics{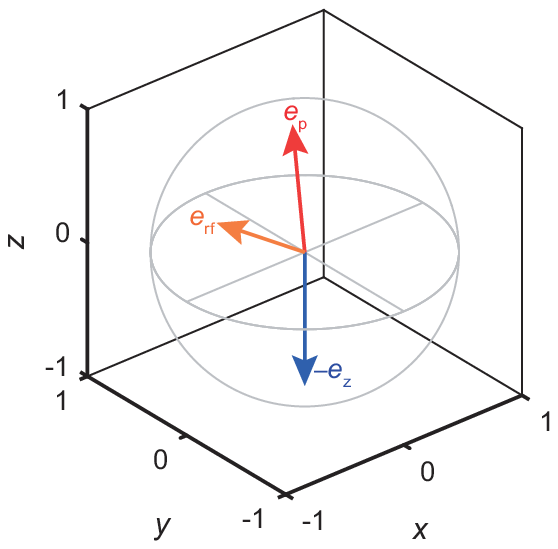}
\caption{
$\bm{\nu}_{\mathrm{n}}(s < -t_2)$ = $-\bm{e}_z$ (blue arrow), $\bm{e}_{\mathrm{p}}$ (red arrow), and $\bm{e}_{\mathrm{rf}}$ (orange arrow) in the sensor coordinate. 
\label{fig_s7}}
\end{center}
\end{figure*}

A primary task in this section is to determine $\phi_{\mathrm{n}}(0)$ accurately.
It should be noted that in real space no sooner does the nuclear spin feel the RF field at $s$ = $-t_1$ than the nuclear spin Bloch vector $\bm{\nu}_{\mathrm{n}}(s)$ has a transverse component.
We have also seen in Sec.~\ref{sec_3} that the precession axis of the nuclear spin is tilted from the $z$ axis, due to the hyperfine interaction.
Therefore, it is important to carefully examine the dynamics of the target nuclear spin in the time range $s \leq 0$.
To set the stage, we review the experimental sequence of Fig.~\ref{fig_s6}(a) step by step.
See also Fig.~\ref{fig_s7}.
\begin{description}
\item[$s < -t_2$] \mbox{}\\
The target nuclear spin is polarized by PulsePol.
We assume that PolY at $k$ = 3 is used, so that the nuclear spin is initialized into $\bm{\nu}_{\mathrm{n}}(s < -t_2) = -\bm{e}_z$ (Sec.~\ref{sec_6}).

\item[$-t_2 \leq s < -t_1$] \mbox{}\\
A 2-$\mu$s-long chirped microwave $\pi$ pulse is applied to flip the NV spin (Sec.~\ref{sec_2}).
During this time, the NV spin is coherently driven from $|m_S = 0 \rangle$ to $|m_S = -1 \rangle$,
and the precession axis of the nuclear spin tilts from $\bm{e}_z$ to $\bm{e}_{\mathrm{p}}$ concurrently (Sec.~\ref{sec_3}.)
We approximate that the precession axis is $\bm{e}_z$ until the midpoint of the chirped pulse ($s$ = $-t_2$).
It then jumps to $\bm{e}_{\mathrm{p}}$ and is fixed subsequently.
In our experiments, $(t_2 - t_1)$ = 2.288~$\mu$s, which is broken down into 1~$\mu$s of the latter half of the chirped pulse,
0.2~$\mu$s of the trigger time for the following RF pulse, and $t_{\mathrm{delay}}$ = 1.088~$\mu$s (Sec.~\ref{sec_2}).

\item[$-t_1 \leq s < 0$] \mbox{}\\
An RF $\pi/2$ pulse at $f_1$ = 215.6~kHz is applied to tip the nuclear spin.
The pulse length is set as $t_1$ = $22/f_1$ = 102.041~$\mu$s (Sec.~\ref{sec_6}).
We define $\bm{e}_{\mathrm{rf}}$ as 
\begin{equation}
\bm{e}_{\mathrm{rf}} = \frac{ \bm{B}_{\mathrm{c}} }{ B_{\mathrm{c}} } = \left(
\begin{array}{c}
\sin \theta_{\mathrm{c}} \cos \phi_{\mathrm{c}} \\
\sin \theta_{\mathrm{c}} \sin \phi_{\mathrm{c}} \\
\cos \theta_{\mathrm{c}}
\end{array} \right)
\end{equation}
with $\theta_{\mathrm{c}}$ = 55.7$^{\circ}$ and $\phi_{\mathrm{c}}$ = 186.2$^{\circ}$ (Sec.~\ref{sec_2}).
When a cosine wave $\cos (2\pi f_1 t)$ is applied on the coil, the RF magnetic field $\bm{B}_{\mathrm{rf}}$ first points to $\bm{e}_{\mathrm{rf}}$,
and after half the oscillation period it points to $-\bm{e}_{\mathrm{rf}}$;
the coil generates a linearly-polarized RF magnetic field along $\bm{e}_{\mathrm{rf}}$.

\item[$0 \leq s < t_0$] \mbox{}\\
The RF pulse is turned off, and the nuclear spin precesses freely.
We set $t_0$ = 6.872~$\mu$s.

\item[$t_0 \leq s$] \mbox{}\\
The free precession of the nuclear spin is detected by the CP sequence with the $(\pi/2)_{\mathrm{Y}}$ readout pulse (Sec.~\ref{sec_3}).
We experimentally determine the precession frequency $f_{\mathrm{p}}$ and $\phi_0$.
\end{description}

With this setting, let us first consider the simplest case, in which both $\bm{e}_{\mathrm{p}} \parallel \bm{e}_z$ and $f_1$ = $f_{\mathrm{p}}$ are satisfied.
The first condition is justified when $B_0$ is higher than a few hundreds of mT.
For instance, at $B_0$ = 1~T, $\theta_{\mathrm{p}}$ should be suppressed to less than 0.1$^{\circ}$.
We define the waveform of the RF field as
\begin{equation}
\bm{B}_{\mathrm{rf}}(s) = 2 b(s) \cos ( 2\pi f_1 s + \phi_{\mathrm{rf}} ) \, \bm{e}_{\mathrm{rf}},
\end{equation}
with
\begin{equation}
b(s) = \left\{
\begin{array}{ll}
(4t_1)^{-1} & ( -t_1 \leq s \leq 0 ) \\
& \\
0 & (\mathrm{otherwise}).
\end{array} \right.
\end{equation}
Because $\bm{e}_{\mathrm{p}} \parallel \bm{e}_z$, the component of $\bm{B}_{\mathrm{rf}}$ projected onto the $xy$ plane ($\bm{B}_{\mathrm{rf}, \perp}$) only acts to rotate the nuclear spin.
In addition, by invoking the rotating wave approximation, it is sufficient to consider a clockwise-rotating component of $\bm{B}_{\mathrm{rf}, \perp}$, which co-rotates with the nuclear spin.
The component of $\bm{B}_{\mathrm{rf}}$ parallel to the $z$ axis ($\bm{B}_{\mathrm{rf}, \parallel}$) modifies the nuclear precession frequency.
However, such a frequency modulation averages out by setting the RF pulse length as an integer multiple of the oscillation period (Sec.~\ref{sec_6}), and the effect of $\bm{B}_{\mathrm{rf}, \parallel}$ becomes negligible.

$\phi_{\mathrm{n}}(0)$ is evaluated as
\begin{equation}
\phi_{\mathrm{n}}(0) = -\phi_{\mathrm{rf}} +\phi_{\mathrm{c}} - \frac{\pi}{2} = -\phi_{\mathrm{rf}} + 96.2^{\circ} \quad (\bmod \,\, 360^{\circ}).
\label{eq_phi_n_1}
\end{equation}
Here, $-\phi_{\mathrm{rf}} +\phi_{\mathrm{c}}$ is the azimuthal angle of the rotation axis of the nuclear spin (note that $\phi_{\mathrm{rf}}$ changes clockwise),
and by the RF $\pi/2$ pulse the nuclear spin ends up in the direction orthogonal to it (the negative sign in $-\frac{\pi}{2}$ reflects the rotation direction of the nuclear spin).

When $f_1 \neq f_{\mathrm{p}}$, the detuning accumulates as $-2\pi (f_{\mathrm{p}} - f_1) t_1$, relative to $-\phi_{\mathrm{rf}} +\phi_{\mathrm{c}}$.
Again, the negative sign is due to the fact that the nuclear spin precesses clockwise, whereas $\phi$ is defined counter-clockwise.
Experimentally obtained $f_{\mathrm{p}}$ is on average $\bar{f_{\mathrm{p}}}$ = 215.7908~kHz,
so the effect of detuning amounts to $-0.319^{\circ} \times 22 = -7.0^{\circ}$.
$\phi_{\mathrm{n}}(0)$ is evaluated as
\begin{equation}
\phi_{\mathrm{n}}(0) = -\phi_{\mathrm{rf}} +\phi_{\mathrm{c}} - \frac{\pi}{2} -2\pi (\bar{f_{\mathrm{p}}} - f_1) t_1 = -\phi_{\mathrm{rf}} + 89.2^{\circ} \quad (\bmod \,\, 360^{\circ}).
\label{eq_phi_n_2}
\end{equation}

In Fig.~4(b) of the main text, we experimentally obtain $\phi_0$ as
\begin{equation}
\phi_0 = \phi_{\mathrm{rf}} + 334.0^{\circ} \quad (\bmod \,\, 360^{\circ}).
\label{eq_phi_0}
\end{equation}
From Eqs.~(\ref{eq_phi}), (\ref{eq_phi_n_2}), and (\ref{eq_phi_0}), we obtain
\begin{equation}
\phi = (-\phi_{\mathrm{rf}} + 89.2^{\circ}) + (\phi_{\mathrm{rf}} + 334.0^{\circ}) + 180^{\circ} = 243.2^{\circ} \quad (\bmod \,\, 360^{\circ}).
\end{equation}
$\phi$ = 243.2$^{\circ}$ is given as the dashed line in the lower panel of Fig.~4(b) of the main text, and
the accuracy range 243.2$\pm$5.3$^{\circ}$ is shown in the left panel of Fig.~4(c).

\begin{figure*}
\begin{center}
\includegraphics{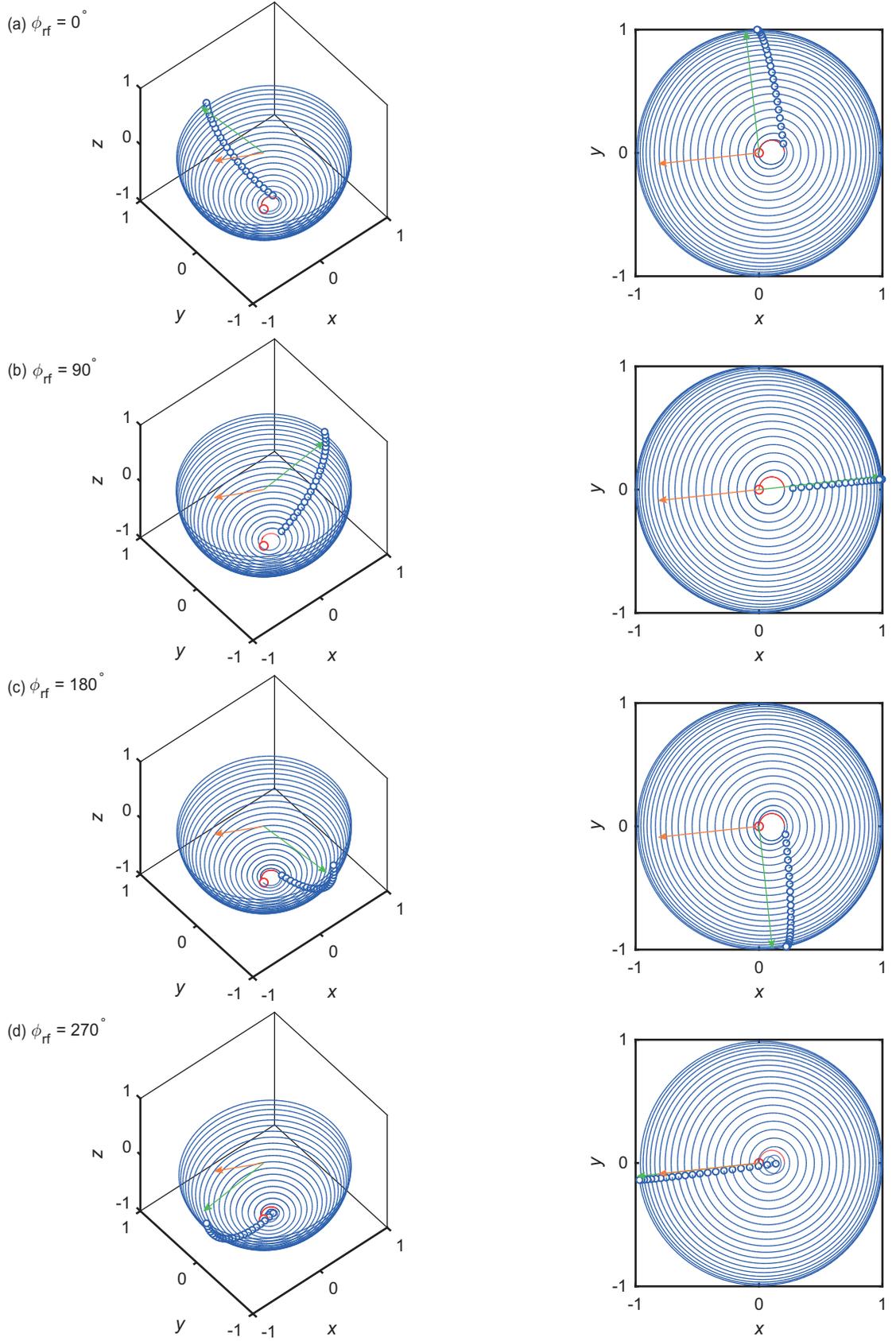}
\caption{
Dynamics of the target nuclear spin with the simulation parameters $\phi$ = 0$^{\circ}$ and $f_1$ = $f_{\mathrm{p}}$ = 215.6~kHz.
(a) $\phi_{\mathrm{rf}}$ = 0$^{\circ}$, (b) 90$^{\circ}$, (c) 180$^{\circ}$, and (d) 270$^{\circ}$.
The orange arrow indicates $\bm{e}_{\mathrm{rf}}$.
The red circles ($\circ$) indicate the initial position.
The red (blue) curves are the trajectories during $-t_2 \leq s < -t_1$ ($-t_1 \leq s \leq 0$).
The blue circles ($\circ$) follow the stroboscopic trajectories at $s$ = $-t_1 + k/f_1$, with $k$ = 0, 1, $\cdots$, 22, similar to the ones observed in the rotating frame.
The green arrows indicate $\phi_{\mathrm{n}}(0)$ evaluated using Eq.~(\ref{eq_phi_n_1}).
The values of simulated $\phi_{\mathrm{n}}(0)$ are
90.9$^{\circ}$ at $\phi_{\mathrm{rf}}$ = 0$^{\circ}$,
4.8$^{\circ}$ at $\phi_{\mathrm{rf}}$ = 90$^{\circ}$,
282.7$^{\circ}$ at $\phi_{\mathrm{rf}}$ = 180$^{\circ}$, and
188.1$^{\circ}$ at $\phi_{\mathrm{rf}}$ = 270$^{\circ}$.
\label{fig_s8}}
\end{center}
\end{figure*}
\begin{figure*}
\begin{center}
\includegraphics{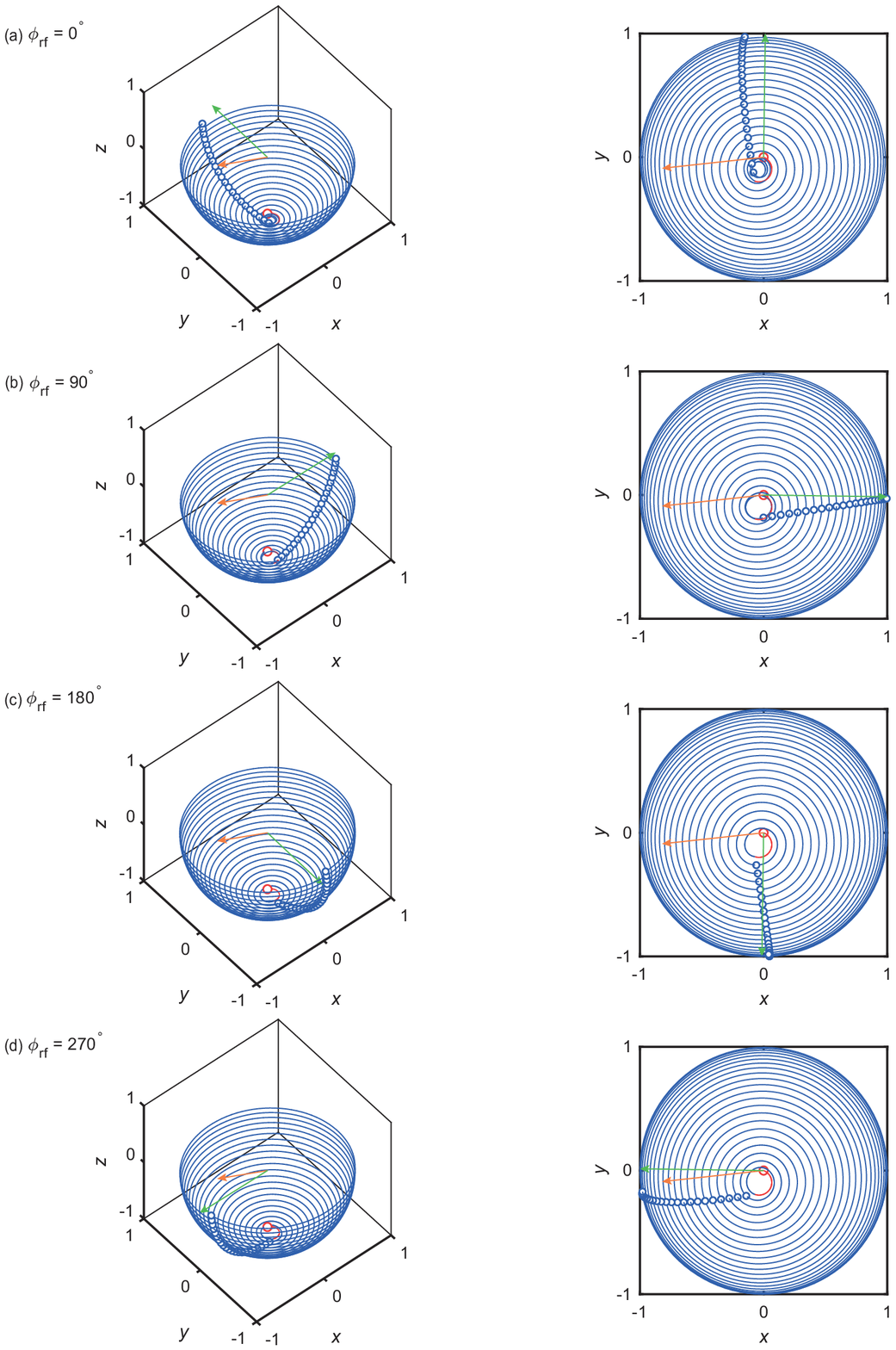}
\caption{
Dynamics of the target nuclear spin with the simulation parameters $\phi$ = 248.8$^{\circ}$, $f_1$ = 215.6~kHz, and $\bar{f_{\mathrm{p}}}$ = 215.7908~kHz.
The green arrows indicate $\phi_{\mathrm{n}}(0)$ evaluated using Eq.~(\ref{eq_phi_n_2}).
The values of simulated $\phi_{\mathrm{n}}(0)$ are
98.9$^{\circ}$ at $\phi_{\mathrm{rf}}$ = 0$^{\circ}$,
358.4$^{\circ}$ at $\phi_{\mathrm{rf}}$ = 90$^{\circ}$,
272.4$^{\circ}$ at $\phi_{\mathrm{rf}}$ = 180$^{\circ}$, and
189.9$^{\circ}$ at $\phi_{\mathrm{rf}}$ = 270$^{\circ}$.
\label{fig_s9}}
\end{center}
\end{figure*}

For more accurate estimation, we consider the case $\bm{e}_{\mathrm{p}} \nparallel \bm{e}_z$.
As mentioned above, this effect is suppressed by applying large $B_0$, but the present experiments are performed at $B_0$ = 36.2~mT,
rendering the polar angle of $\bm{e}_{\mathrm{p}}$ to be 5.9$^{\circ}$ (Sec.~\ref{sec_3}).
The effect of $\bm{e}_{\mathrm{p}}$ is worth a careful analysis.
This is best done by simulating real-space trajectories of the target nuclear spin in the time range $-t_2 \leq s \leq 0$, based on the Bloch equation:
\begin{equation}
\frac{ d\bm{\nu}_{\mathrm{n}} (s) }{ ds } = 2\pi \bm{\nu}_{\mathrm{n}}(s) \times
[ f_{\mathrm{p}} \, \bm{e}_{\mathrm{p}} + B_{\mathrm{rf}}(s) \, \bm{e}_{\mathrm{rf}} ],
\label{eq_bloch}
\end{equation}
where $b(s)$ appearing in $B_{\mathrm{rf}}(s)$ is now
\begin{equation}
b(s) = \left\{
\begin{array}{ll}
(4 t_1 | \bm{e}_{\mathrm{rf}} \times \bm{e}_{\mathrm{p}} |)^{-1} & ( -t_1 \leq s \leq 0 ) \\
& \\
0 & (\mathrm{otherwise}).
\end{array} \right.
\end{equation}

To gain physical insights, we first perform simulations by setting $\phi$ = 0$^{\circ}$ and $f_1$ = $f_{\mathrm{p}}$ = 215.6~kHz.
The results are shown in Fig.~\ref{fig_s8}.
Simulated $\phi_{\mathrm{n}}(0)$ deviates from Eq.~(\ref{eq_phi_n_1}) on the order of $\theta_{\mathrm{p}}$.
We observe that the inital precession about $\bm{e}_{\mathrm{p}}$, before the RF pulse is applied, is a major source of deviation.
Since the azimuthal angle of $\bm{e}_{\mathrm{p}}$ is $\phi$, this type of deviation sinusoidally depends on $\phi$ as well as on $\phi_{\mathrm{rf}}$ and $(t_2 - t_1)$.
One way to suppress this effect at low fields may be to set $(t_2 - t_1)$ as an integer multiple of the precession period,
so that the nuclear spin returns to $-\bm{e}_z$ when the RF field is applied.

We remark that there is a more subtle, additional effect of $\bm{e}_{\mathrm{p}}$ that is in play during the RF pulse.
Once the RF field is applied, the nuclear spin rotates, to a good approximation, around $\bm{B}_{\mathrm{rf}, \perp}$.
However, because now the nuclear spin feels a circularly-polarized RF field that is rotating in the plane perpendicular to $\bm{e}_{\mathrm{p}}$,
the nuclear spin does not end up in the $xy$ plane exactly (even when the nuclear spin starts to rotate from $-\bm{e}_z$).
When the nuclear spin is projected onto the $xy$ plane, this effect is seen to depend on $\phi$, but not on $\phi_{\mathrm{rf}}$.
What is seen in Fig.~\ref{fig_s8} is a cumulative effect of two roles that $\bm{e}_{\mathrm{p}}$ plays,
making the deviation from Eq.~(\ref{eq_phi_n_1}) less systematic as changing $\phi_{\mathrm{rf}}$.
If $(t_2 - t_1)$ is as an integer multiple of the precession period, only the latter is effective.
The deviation becomes systematic, and the analysis will be facilitated.

We set $\phi$ and $\phi_{\mathrm{rf}}$ as parameters in the simulation, and find the value of $\phi$ that best reproduces the experimental data,
The precession frequency is set as $\bar{f_{\mathrm{p}}}$, and assumed to be independent of $\phi_{\mathrm{rf}}$.
The simulation result for $\phi$ = 248.8$^{\circ}$ is shown in Fig.~\ref{fig_s9}, for which the deviation from the experimental data is minimized.
$\phi$ = 248.8$^{\circ}$ is given as the dotted line in the lower panel of Fig.~4(b) of the main text, and
the accuracy range 248.8$\pm$2.7$^{\circ}$ is shown in the middle panel of Fig.~4(c).

Strictly speaking, measured precession frequencies $f_{\mathrm{p}}$ vary from one measurement to another.
This is partly attributed to errors in the fits, but it is also conceivable that the precession frequency indeed differs in different measurements,
due to, for instance, temperature drifts (which can change $B_0$ provided by a permanent magnet).
Therefore, lastly, we perform simulations by setting $\phi$ as the only parameter and using the values of $f_{\mathrm{p}}$ for the respective runs.
The results are shown as the circles ($\circ$) in the lower panel of Fig.~4(b) of the main text, and their average value, 247.8$^{\circ}$, is given as the solid line.
The accuracy range 247.8$\pm$4.1$^{\circ}$ is shown in the right panel of Fig.~4(c).

The lattice site that falls on these accuracy ranges are shown as blue circles in Fig.~4(c), which have $\phi$ = 250.9$^{\circ}$.
Therefore, we determine is the position of the target nuclear spin as ($r$, $\theta$, $\phi$) = (6.84~\AA, 94.8$^{\circ}$, 250.9$^{\circ}$).

\section{\uppercase{D}\lowercase{emonstration of the protocol on a different single $^{13}$}\uppercase{C}\lowercase{ nuclear spin}\label{sec_10}}
\begin{figure*}
\begin{center}
\includegraphics{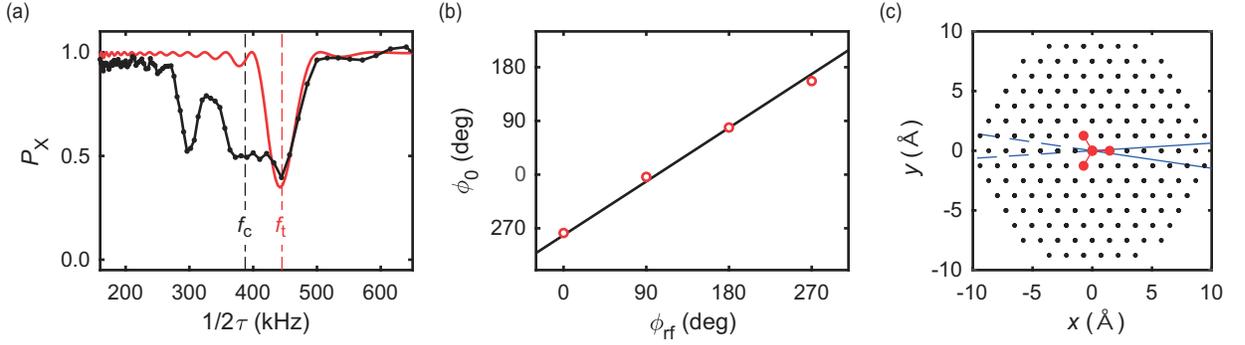}
\caption{
(a) Data in Fig.~2(a) of the main text, with a simulation to a different dip at $f_{\mathrm{t}}$ = 443.4~kHz.
(b) $\phi_{\mathrm{rf}}$ vs. $\phi_0$ for the second target nuclear spin.
The experimental parameters are $f_1$ = 503.0~kHz, $t_0$ = 4.403~$\mu$s, $t_1$ = 214.725~$\mu$s (108 periods), and $t_2$ = 216.797~$\mu$s.
$\bar{f_{\mathrm{p}}}$ = 503.0~kHz is obtained.
(c) Estimated $\phi$ of the second target nuclear spin.
The accuracy ranges are 357.7$\pm$8.6$^{\circ}$ for $0 \leq \theta < \frac{\pi}{2}$ and 177.7$\pm$8.6$^{\circ}$ for $\frac{\pi}{2} \leq \theta < \pi$.
\label{fig_s10}}
\end{center}
\end{figure*}

Here, we show results on a single nuclear spin different from the one discussed in the main text.
Figure~\ref{fig_s10}(a) shows the NMR spectrum given in Fig.~2(a) of the main text.
We observe that there is another dip at $f_{\mathrm{t}}$ = 443.4~kHz, albeit overlapping with the broad dip arising from the bath nuclei.
We examined this signal in more detail (as done on the first nuclear spin in the main text), and determined its hyperfine parameters as $A_{\parallel}$ = 112.1~kHz and $A_{\perp}$ = 59.9~kHz.
Unfortunately, we have not been able to narrow down the lattice site it belongs to, because we do not find theoretical values sufficiently close to them.
Although we do not have $r$ and $\theta$ of this second nuclear spin, the protocol can still be applied to estimate $\phi$.
A summary of the experiments are shown in Figs.~\ref{fig_s10}(b) and (c).
All the experimental parameters, given in the caption of Fig.~\ref{fig_s10}, are optimized for this nuclear spin.
Note that there remain two possible ranges of $\phi$, differing by 180$^{\circ}$, due to the lack of knowledge on $\theta$.
Nonetheless, these values are different from those obtained from the first target nuclear spin, and support our claim that the values of $\phi$ obtained in our protocol are specific to the individual nuclear spins.

\section{\uppercase{O}\lowercase{bservation of bath $^{13}$}\uppercase{C}\lowercase{ nuclear spins by coherent averaging}\label{sec_11}}
\begin{figure*}
\begin{center}
\includegraphics{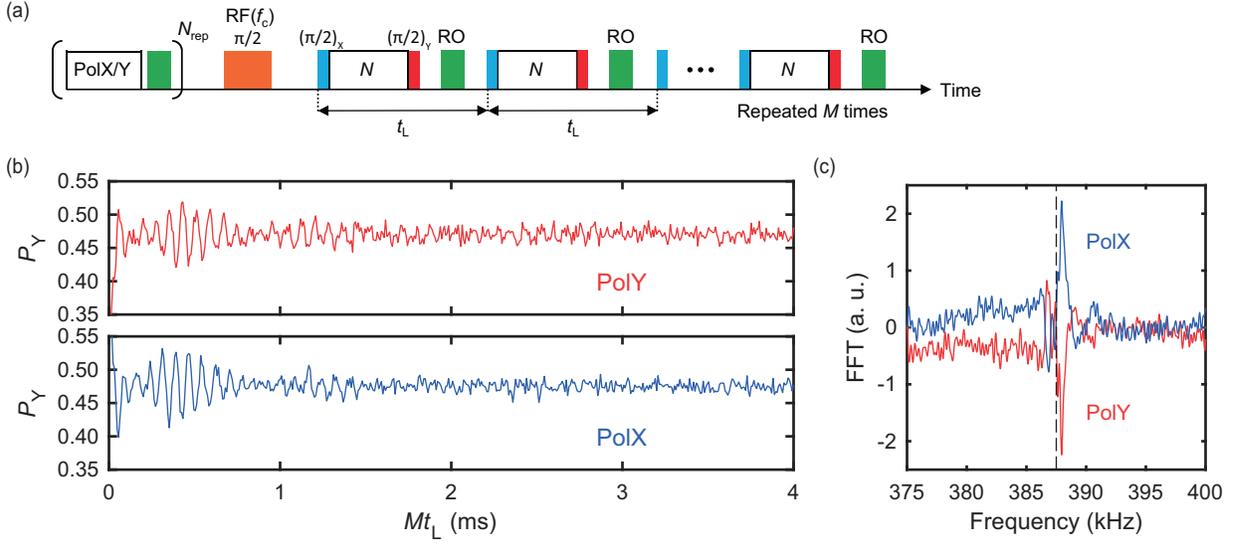}
\caption{
(a) Pulse protocol to observe the bath nuclear spins.
The RF frequency is set at $f_{\mathrm{c}} = \gamma_{\mathrm{c}} B_0$ = 387.5~kHz.
(b) $P_{\mathrm{Y}}$ as a function of $M t_{\mathrm{L}}$.
(c) FFT of (b).
The dashed line indicates $f_{\mathrm{c}}$.
\label{fig_s11}}
\end{center}
\end{figure*}

As commented in the main text, our protocol can be combined with the high-resolution spectroscopy method reported in~\cite{SGS+17s,AS18s,GBL+18s,BCZD17s}.
We reinforce this claim by observing bath $^{13}$C nuclear spins, using the sequence shown in Fig.~\ref{fig_s11}(a).
On the one hand, this sequence is essentially the same as coherently averaged synchronized readout described in~\cite{GBL+18s};
after inducing free precessions of bath nuclei by an RF $\pi/2$ pulse, we record them by repeating the CP sequence with the regular interval of $t_{\mathrm{L}}$.
On the other hand, coherent averaging of the signal from a handful of nuclear spins is possible only when PulsePol is applied prior to the RF $\pi/2$ pulse.
Even though the bath nuclei contain multiple nuclear spins, they cannot be regarded as an ensemble as in the case of external proton spins in~\cite{GBL+18s}.
Rather, the bath nuclei are a collection of independent single nuclear spins, the individual hyperfine parameters of which could in principle be resolved by high-resolution spectroscopy.

We polarize the bath nuclei by PulsePol with 2$\tau_{\mathrm{pol}}$ = 7.744~$\mu$s (= $3f_{\mathrm{c}}^{-1}$), $N_{\mathrm{pol}}$ = 1, and $N_{\mathrm{rep}}$ = 200.
The RF $\pi/2$ pulse tuned at $f_{\mathrm{c}}$ is applied, and the CP sequence with $N$ = 2 and 2$\tau$ = 2.581~$\mu$s (= $f_{\mathrm{c}}^{-1}$)
was repeated $M$ = 501 times with the interval of $t_{\mathrm{L}}$ = 8.000~$\mu$s.
To mitigate possible back actions on the nuclear spins, the number of pulses in the CP sequence was minimized ($N$ = 2).
For the same reason, the signals are undersampled at $m$ = 6.
Figure~\ref{fig_s11}(b) shows measured oscillations for PolY and PolX, and their FFT spectra (real part, with the frequency shift due to undersampling corrected) are shown in Fig.~\ref{fig_s11}(c).
Clearly, the oscillation phases are opposite, consistent with our main result on the single nuclear spin [Fig.~4(b) of the main text].
While a further analysis of Figs.~\ref{fig_s11}(b) and (c) is outside of the scope of the present work,
multiple signals present in the data suggest that we are detecting multiple nuclear spins simultaneously
and the phases of the respective frequency components carry the information on the azimuthal angles of the respective nuclear spins.
We note that the sequence in Fig.~\ref{fig_s11}(a) does not use a microwave $\pi$ pulse prior to the RF $\pi/2$ pulse,
and therefore the precession axis of the nuclear spins is fixed as $\bm{e}_z$.

Finally, we checked that the signal decay time does not depend on $t_{\mathrm{L}}$,
indicating that the present decay time is not limited by the back actions that the sequence exerts on the nuclear spins.
An understanding of sources of the decay will be a subject of future research.


\begin{thebibliography}{39}%
\makeatletter
\providecommand \@ifxundefined [1]{%
 \@ifx{#1\undefined}
}%
\providecommand \@ifnum [1]{%
 \ifnum #1\expandafter \@firstoftwo
 \else \expandafter \@secondoftwo
 \fi
}%
\providecommand \@ifx [1]{%
 \ifx #1\expandafter \@firstoftwo
 \else \expandafter \@secondoftwo
 \fi
}%
\providecommand \natexlab [1]{#1}%
\providecommand \enquote  [1]{``#1''}%
\providecommand \bibnamefont  [1]{#1}%
\providecommand \bibfnamefont [1]{#1}%
\providecommand \citenamefont [1]{#1}%
\providecommand \href@noop [0]{\@secondoftwo}%
\providecommand \href [0]{\begingroup \@sanitize@url \@href}%
\providecommand \@href[1]{\@@startlink{#1}\@@href}%
\providecommand \@@href[1]{\endgroup#1\@@endlink}%
\providecommand \@sanitize@url [0]{\catcode `\\12\catcode `\$12\catcode
  `\&12\catcode `\#12\catcode `\^12\catcode `\_12\catcode `\%12\relax}%
\providecommand \@@startlink[1]{}%
\providecommand \@@endlink[0]{}%
\providecommand \url  [0]{\begingroup\@sanitize@url \@url }%
\providecommand \@url [1]{\endgroup\@href {#1}{\urlprefix }}%
\providecommand \urlprefix  [0]{URL }%
\providecommand \Eprint [0]{\href }%
\providecommand \doibase [0]{http://dx.doi.org/}%
\providecommand \selectlanguage [0]{\@gobble}%
\providecommand \bibinfo  [0]{\@secondoftwo}%
\providecommand \bibfield  [0]{\@secondoftwo}%
\providecommand \translation [1]{[#1]}%
\providecommand \BibitemOpen [0]{}%
\providecommand \bibitemStop [0]{}%
\providecommand \bibitemNoStop [0]{.\EOS\space}%
\providecommand \EOS [0]{\spacefactor3000\relax}%
\providecommand \BibitemShut  [1]{\csname bibitem#1\endcsname}%
\let\auto@bib@innerbib\@empty
\bibitem [{\citenamefont {Schirhagl}\ \emph {et~al.}(2014)\citenamefont
  {Schirhagl}, \citenamefont {Chang}, \citenamefont {Loretz},\ and\
  \citenamefont {Degen}}]{SCLD14}%
  \BibitemOpen
  \bibfield  {author} {\bibinfo {author} {\bibfnamefont {R.}~\bibnamefont
  {Schirhagl}}, \bibinfo {author} {\bibfnamefont {K.}~\bibnamefont {Chang}},
  \bibinfo {author} {\bibfnamefont {M.}~\bibnamefont {Loretz}}, \ and\ \bibinfo
  {author} {\bibfnamefont {C.~L.}\ \bibnamefont {Degen}},\ }\href@noop {}
  {\bibfield  {journal} {\bibinfo  {journal} {Annu.\ Rev.\ Phys.\ Chem.}\
  }\textbf {\bibinfo {volume} {65}},\ \bibinfo {pages} {83} (\bibinfo {year}
  {2014})}\BibitemShut {NoStop}%
\bibitem [{\citenamefont {Rondin}\ \emph {et~al.}(2014)\citenamefont {Rondin},
  \citenamefont {Tetienne}, \citenamefont {Hingant}, \citenamefont {Roch},
  \citenamefont {Maletinsky},\ and\ \citenamefont {Jacques}}]{RTH+14}%
  \BibitemOpen
  \bibfield  {author} {\bibinfo {author} {\bibfnamefont {L.}~\bibnamefont
  {Rondin}}, \bibinfo {author} {\bibfnamefont {J.-P.}\ \bibnamefont
  {Tetienne}}, \bibinfo {author} {\bibfnamefont {T.}~\bibnamefont {Hingant}},
  \bibinfo {author} {\bibfnamefont {J.-F.}\ \bibnamefont {Roch}}, \bibinfo
  {author} {\bibfnamefont {P.}~\bibnamefont {Maletinsky}}, \ and\ \bibinfo
  {author} {\bibfnamefont {V.}~\bibnamefont {Jacques}},\ }\href@noop {}
  {\bibfield  {journal} {\bibinfo  {journal} {Rep.\ Prog.\ Phys.}\ }\textbf
  {\bibinfo {volume} {77}},\ \bibinfo {pages} {056503} (\bibinfo {year}
  {2014})}\BibitemShut {NoStop}%
\bibitem [{\citenamefont {Abe}\ and\ \citenamefont {Sasaki}(2018)}]{AS18}%
  \BibitemOpen
  \bibfield  {author} {\bibinfo {author} {\bibfnamefont {E.}~\bibnamefont
  {Abe}}\ and\ \bibinfo {author} {\bibfnamefont {K.}~\bibnamefont {Sasaki}},\
  }\href@noop {} {\bibfield  {journal} {\bibinfo  {journal} {J.\ Appl.\ Phys.}\
  }\textbf {\bibinfo {volume} {123}},\ \bibinfo {pages} {161101} (\bibinfo
  {year} {2018})}\BibitemShut {NoStop}%
\bibitem [{\citenamefont {Mamin}\ \emph {et~al.}(2013)\citenamefont {Mamin},
  \citenamefont {Kim}, \citenamefont {Sherwood}, \citenamefont {Rettner},
  \citenamefont {Ohno}, \citenamefont {Awschalom},\ and\ \citenamefont
  {Rugar}}]{MKS+13}%
  \BibitemOpen
  \bibfield  {author} {\bibinfo {author} {\bibfnamefont {H.~J.}\ \bibnamefont
  {Mamin}}, \bibinfo {author} {\bibfnamefont {M.}~\bibnamefont {Kim}}, \bibinfo
  {author} {\bibfnamefont {M.~H.}\ \bibnamefont {Sherwood}}, \bibinfo {author}
  {\bibfnamefont {C.~T.}\ \bibnamefont {Rettner}}, \bibinfo {author}
  {\bibfnamefont {K.}~\bibnamefont {Ohno}}, \bibinfo {author} {\bibfnamefont
  {D.~D.}\ \bibnamefont {Awschalom}}, \ and\ \bibinfo {author} {\bibfnamefont
  {D.}~\bibnamefont {Rugar}},\ }\href@noop {} {\bibfield  {journal} {\bibinfo
  {journal} {Science}\ }\textbf {\bibinfo {volume} {339}},\ \bibinfo {pages}
  {557} (\bibinfo {year} {2013})}\BibitemShut {NoStop}%
\bibitem [{\citenamefont {Staudacher}\ \emph {et~al.}(2013)\citenamefont
  {Staudacher}, \citenamefont {Shi}, \citenamefont {Pezzagna}, \citenamefont
  {Meijer}, \citenamefont {Du}, \citenamefont {Meriles}, \citenamefont
  {Reinhard},\ and\ \citenamefont {Wrachtrup}}]{SSP+13}%
  \BibitemOpen
  \bibfield  {author} {\bibinfo {author} {\bibfnamefont {T.}~\bibnamefont
  {Staudacher}}, \bibinfo {author} {\bibfnamefont {F.}~\bibnamefont {Shi}},
  \bibinfo {author} {\bibfnamefont {S.}~\bibnamefont {Pezzagna}}, \bibinfo
  {author} {\bibfnamefont {J.}~\bibnamefont {Meijer}}, \bibinfo {author}
  {\bibfnamefont {J.}~\bibnamefont {Du}}, \bibinfo {author} {\bibfnamefont
  {C.~A.}\ \bibnamefont {Meriles}}, \bibinfo {author} {\bibfnamefont
  {F.}~\bibnamefont {Reinhard}}, \ and\ \bibinfo {author} {\bibfnamefont
  {J.}~\bibnamefont {Wrachtrup}},\ }\href@noop {} {\bibfield  {journal}
  {\bibinfo  {journal} {Science}\ }\textbf {\bibinfo {volume} {339}},\ \bibinfo
  {pages} {561} (\bibinfo {year} {2013})}\BibitemShut {NoStop}%
\bibitem [{\citenamefont {Loretz}\ \emph {et~al.}(2014)\citenamefont {Loretz},
  \citenamefont {Pezzagna}, \citenamefont {Meijer},\ and\ \citenamefont
  {Degen}}]{LPMD14}%
  \BibitemOpen
  \bibfield  {author} {\bibinfo {author} {\bibfnamefont {M.}~\bibnamefont
  {Loretz}}, \bibinfo {author} {\bibfnamefont {S.}~\bibnamefont {Pezzagna}},
  \bibinfo {author} {\bibfnamefont {J.}~\bibnamefont {Meijer}}, \ and\ \bibinfo
  {author} {\bibfnamefont {C.~L.}\ \bibnamefont {Degen}},\ }\href@noop {}
  {\bibfield  {journal} {\bibinfo  {journal} {Appl.\ Phys.\ Lett.}\ }\textbf
  {\bibinfo {volume} {104}},\ \bibinfo {pages} {033102} (\bibinfo {year}
  {2014})}\BibitemShut {NoStop}%
\bibitem [{\citenamefont {M{\"u}ller}\ \emph {et~al.}(2014)\citenamefont
  {M{\"u}ller}, \citenamefont {Kong}, \citenamefont {Cai}, \citenamefont
  {Melentijevi{\'c}}, \citenamefont {Stacey}, \citenamefont {Markham},
  \citenamefont {Twitchen}, \citenamefont {Isoya}, \citenamefont {Pezzagna},
  \citenamefont {Meijer}, \citenamefont {Du}, \citenamefont {Plenio},
  \citenamefont {Naydenov}, \citenamefont {McGuinness},\ and\ \citenamefont
  {Jelezko}}]{MKC+14}%
  \BibitemOpen
  \bibfield  {author} {\bibinfo {author} {\bibfnamefont {C.}~\bibnamefont
  {M{\"u}ller}}, \bibinfo {author} {\bibfnamefont {X.}~\bibnamefont {Kong}},
  \bibinfo {author} {\bibfnamefont {J.-M.}\ \bibnamefont {Cai}}, \bibinfo
  {author} {\bibfnamefont {K.}~\bibnamefont {Melentijevi{\'c}}}, \bibinfo
  {author} {\bibfnamefont {A.}~\bibnamefont {Stacey}}, \bibinfo {author}
  {\bibfnamefont {M.}~\bibnamefont {Markham}}, \bibinfo {author} {\bibfnamefont
  {D.}~\bibnamefont {Twitchen}}, \bibinfo {author} {\bibfnamefont
  {J.}~\bibnamefont {Isoya}}, \bibinfo {author} {\bibfnamefont
  {S.}~\bibnamefont {Pezzagna}}, \bibinfo {author} {\bibfnamefont
  {J.}~\bibnamefont {Meijer}}, \bibinfo {author} {\bibfnamefont {J.~F.}\
  \bibnamefont {Du}}, \bibinfo {author} {\bibfnamefont {M.~B.}\ \bibnamefont
  {Plenio}}, \bibinfo {author} {\bibfnamefont {B.}~\bibnamefont {Naydenov}},
  \bibinfo {author} {\bibfnamefont {L.~P.}\ \bibnamefont {McGuinness}}, \ and\
  \bibinfo {author} {\bibfnamefont {F.}~\bibnamefont {Jelezko}},\ }\href@noop
  {} {\bibfield  {journal} {\bibinfo  {journal} {Nat.\ Commun.}\ }\textbf
  {\bibinfo {volume} {5}},\ \bibinfo {pages} {4703} (\bibinfo {year}
  {2014})}\BibitemShut {NoStop}%
\bibitem [{\citenamefont {H{\"a}berle}\ \emph {et~al.}(2015)\citenamefont
  {H{\"a}berle}, \citenamefont {Schmid-Lorch}, \citenamefont {Reinhard},\ and\
  \citenamefont {Wrachtrup}}]{HSR+15}%
  \BibitemOpen
  \bibfield  {author} {\bibinfo {author} {\bibfnamefont {T.}~\bibnamefont
  {H{\"a}berle}}, \bibinfo {author} {\bibfnamefont {D.}~\bibnamefont
  {Schmid-Lorch}}, \bibinfo {author} {\bibfnamefont {F.}~\bibnamefont
  {Reinhard}}, \ and\ \bibinfo {author} {\bibfnamefont {J.}~\bibnamefont
  {Wrachtrup}},\ }\href@noop {} {\bibfield  {journal} {\bibinfo  {journal}
  {Nat.\ Nanotechnol.}\ }\textbf {\bibinfo {volume} {10}},\ \bibinfo {pages}
  {125} (\bibinfo {year} {2015})}\BibitemShut {NoStop}%
\bibitem [{\citenamefont {DeVience}\ \emph {et~al.}(2015)\citenamefont
  {DeVience}, \citenamefont {Pham}, \citenamefont {Lovchinsky}, \citenamefont
  {Sushkov}, \citenamefont {Bar-Gill}, \citenamefont {Belthangady},
  \citenamefont {Casola}, \citenamefont {Corbett}, \citenamefont {Zhang},
  \citenamefont {Lukin}, \citenamefont {Park}, \citenamefont {Yacoby},\ and\
  \citenamefont {Walsworth}}]{DPL+15}%
  \BibitemOpen
  \bibfield  {author} {\bibinfo {author} {\bibfnamefont {S.~J.}\ \bibnamefont
  {DeVience}}, \bibinfo {author} {\bibfnamefont {L.~M.}\ \bibnamefont {Pham}},
  \bibinfo {author} {\bibfnamefont {I.}~\bibnamefont {Lovchinsky}}, \bibinfo
  {author} {\bibfnamefont {A.~O.}\ \bibnamefont {Sushkov}}, \bibinfo {author}
  {\bibfnamefont {N.}~\bibnamefont {Bar-Gill}}, \bibinfo {author}
  {\bibfnamefont {C.}~\bibnamefont {Belthangady}}, \bibinfo {author}
  {\bibfnamefont {F.}~\bibnamefont {Casola}}, \bibinfo {author} {\bibfnamefont
  {M.}~\bibnamefont {Corbett}}, \bibinfo {author} {\bibfnamefont
  {H.}~\bibnamefont {Zhang}}, \bibinfo {author} {\bibfnamefont
  {M.}~\bibnamefont {Lukin}}, \bibinfo {author} {\bibfnamefont
  {H.}~\bibnamefont {Park}}, \bibinfo {author} {\bibfnamefont {A.}~\bibnamefont
  {Yacoby}}, \ and\ \bibinfo {author} {\bibfnamefont {R.~L.}\ \bibnamefont
  {Walsworth}},\ }\href@noop {} {\bibfield  {journal} {\bibinfo  {journal}
  {Nat.\ Nanotechnol.}\ }\textbf {\bibinfo {volume} {10}},\ \bibinfo {pages}
  {129} (\bibinfo {year} {2015})}\BibitemShut {NoStop}%
\bibitem [{\citenamefont {Sushkov}\ \emph {et~al.}(2014)\citenamefont
  {Sushkov}, \citenamefont {Lovchinsky}, \citenamefont {Chisholm},
  \citenamefont {Walsworth}, \citenamefont {Park},\ and\ \citenamefont
  {Lukin}}]{SLC+14}%
  \BibitemOpen
  \bibfield  {author} {\bibinfo {author} {\bibfnamefont {A.~O.}\ \bibnamefont
  {Sushkov}}, \bibinfo {author} {\bibfnamefont {I.}~\bibnamefont {Lovchinsky}},
  \bibinfo {author} {\bibfnamefont {N.}~\bibnamefont {Chisholm}}, \bibinfo
  {author} {\bibfnamefont {R.~L.}\ \bibnamefont {Walsworth}}, \bibinfo {author}
  {\bibfnamefont {H.}~\bibnamefont {Park}}, \ and\ \bibinfo {author}
  {\bibfnamefont {M.~D.}\ \bibnamefont {Lukin}},\ }\href@noop {} {\bibfield
  {journal} {\bibinfo  {journal} {Phys.\ Rev.\ Lett.}\ }\textbf {\bibinfo
  {volume} {113}},\ \bibinfo {pages} {197601} (\bibinfo {year}
  {2014})}\BibitemShut {NoStop}%
\bibitem [{\citenamefont {Lovchinsky}\ \emph {et~al.}(2016)\citenamefont
  {Lovchinsky}, \citenamefont {Sushkov}, \citenamefont {Urbach}, \citenamefont
  {de~Leon}, \citenamefont {Choi}, \citenamefont {De~Greve}, \citenamefont
  {Evans}, \citenamefont {Gertner}, \citenamefont {Bersin}, \citenamefont
  {M{\"u}ller}, \citenamefont {McGuinness}, \citenamefont {Jelezko},
  \citenamefont {Walsworth}, \citenamefont {Park},\ and\ \citenamefont
  {Lukin}}]{LSU+16}%
  \BibitemOpen
  \bibfield  {author} {\bibinfo {author} {\bibfnamefont {I.}~\bibnamefont
  {Lovchinsky}}, \bibinfo {author} {\bibfnamefont {A.~O.}\ \bibnamefont
  {Sushkov}}, \bibinfo {author} {\bibfnamefont {E.}~\bibnamefont {Urbach}},
  \bibinfo {author} {\bibfnamefont {N.~P.}\ \bibnamefont {de~Leon}}, \bibinfo
  {author} {\bibfnamefont {S.}~\bibnamefont {Choi}}, \bibinfo {author}
  {\bibfnamefont {K.}~\bibnamefont {De~Greve}}, \bibinfo {author}
  {\bibfnamefont {R.}~\bibnamefont {Evans}}, \bibinfo {author} {\bibfnamefont
  {R.}~\bibnamefont {Gertner}}, \bibinfo {author} {\bibfnamefont
  {E.}~\bibnamefont {Bersin}}, \bibinfo {author} {\bibfnamefont
  {C.}~\bibnamefont {M{\"u}ller}}, \bibinfo {author} {\bibfnamefont
  {L.}~\bibnamefont {McGuinness}}, \bibinfo {author} {\bibfnamefont
  {F.}~\bibnamefont {Jelezko}}, \bibinfo {author} {\bibfnamefont {R.~L.}\
  \bibnamefont {Walsworth}}, \bibinfo {author} {\bibfnamefont {H.}~\bibnamefont
  {Park}}, \ and\ \bibinfo {author} {\bibfnamefont {M.~D.}\ \bibnamefont
  {Lukin}},\ }\href@noop {} {\bibfield  {journal} {\bibinfo  {journal}
  {Science}\ }\textbf {\bibinfo {volume} {351}},\ \bibinfo {pages} {836}
  (\bibinfo {year} {2016})}\BibitemShut {NoStop}%
\bibitem [{\citenamefont {Aslam}\ \emph {et~al.}(2017)\citenamefont {Aslam},
  \citenamefont {Pfender}, \citenamefont {Neumann}, \citenamefont {Reuter},
  \citenamefont {Zappe}, \citenamefont {de~Oliveira}, \citenamefont
  {Denisenko}, \citenamefont {Sumiya}, \citenamefont {Onoda}, \citenamefont
  {Isoya},\ and\ \citenamefont {Wrachtrup}}]{APN+17}%
  \BibitemOpen
  \bibfield  {author} {\bibinfo {author} {\bibfnamefont {N.}~\bibnamefont
  {Aslam}}, \bibinfo {author} {\bibfnamefont {M.}~\bibnamefont {Pfender}},
  \bibinfo {author} {\bibfnamefont {P.}~\bibnamefont {Neumann}}, \bibinfo
  {author} {\bibfnamefont {R.}~\bibnamefont {Reuter}}, \bibinfo {author}
  {\bibfnamefont {A.}~\bibnamefont {Zappe}}, \bibinfo {author} {\bibfnamefont
  {F.~F.}\ \bibnamefont {de~Oliveira}}, \bibinfo {author} {\bibfnamefont
  {A.}~\bibnamefont {Denisenko}}, \bibinfo {author} {\bibfnamefont
  {H.}~\bibnamefont {Sumiya}}, \bibinfo {author} {\bibfnamefont
  {S.}~\bibnamefont {Onoda}}, \bibinfo {author} {\bibfnamefont
  {J.}~\bibnamefont {Isoya}}, \ and\ \bibinfo {author} {\bibfnamefont
  {J.}~\bibnamefont {Wrachtrup}},\ }\href@noop {} {\bibfield  {journal}
  {\bibinfo  {journal} {Science}\ }\textbf {\bibinfo {volume} {357}},\ \bibinfo
  {pages} {67} (\bibinfo {year} {2017})}\BibitemShut {NoStop}%
\bibitem [{\citenamefont {Glenn}\ \emph {et~al.}(2018)\citenamefont {Glenn},
  \citenamefont {Bucher}, \citenamefont {Lee}, \citenamefont {Lukin},
  \citenamefont {Park},\ and\ \citenamefont {Walsworth}}]{GBL+18}%
  \BibitemOpen
  \bibfield  {author} {\bibinfo {author} {\bibfnamefont {D.~R.}\ \bibnamefont
  {Glenn}}, \bibinfo {author} {\bibfnamefont {D.~B.}\ \bibnamefont {Bucher}},
  \bibinfo {author} {\bibfnamefont {J.}~\bibnamefont {Lee}}, \bibinfo {author}
  {\bibfnamefont {M.~D.}\ \bibnamefont {Lukin}}, \bibinfo {author}
  {\bibfnamefont {H.}~\bibnamefont {Park}}, \ and\ \bibinfo {author}
  {\bibfnamefont {R.~L.}\ \bibnamefont {Walsworth}},\ }\href@noop {} {\bibfield
   {journal} {\bibinfo  {journal} {Nature}\ }\textbf {\bibinfo {volume}
  {555}},\ \bibinfo {pages} {351} (\bibinfo {year} {2018})}\BibitemShut
  {NoStop}%
\bibitem [{\citenamefont {Staudacher}\ \emph {et~al.}(2015)\citenamefont
  {Staudacher}, \citenamefont {Raatz}, \citenamefont {Pezzagna}, \citenamefont
  {Meijer}, \citenamefont {Reinhard}, \citenamefont {Meriles},\ and\
  \citenamefont {Wrachtrup}}]{SRP+15}%
  \BibitemOpen
  \bibfield  {author} {\bibinfo {author} {\bibfnamefont {T.}~\bibnamefont
  {Staudacher}}, \bibinfo {author} {\bibfnamefont {N.}~\bibnamefont {Raatz}},
  \bibinfo {author} {\bibfnamefont {S.}~\bibnamefont {Pezzagna}}, \bibinfo
  {author} {\bibfnamefont {J.}~\bibnamefont {Meijer}}, \bibinfo {author}
  {\bibfnamefont {F.}~\bibnamefont {Reinhard}}, \bibinfo {author}
  {\bibfnamefont {C.~A.}\ \bibnamefont {Meriles}}, \ and\ \bibinfo {author}
  {\bibfnamefont {J.}~\bibnamefont {Wrachtrup}},\ }\href@noop {} {\bibfield
  {journal} {\bibinfo  {journal} {Nat.\ Commun.}\ }\textbf {\bibinfo {volume}
  {6}},\ \bibinfo {pages} {8527} (\bibinfo {year} {2015})}\BibitemShut
  {NoStop}%
\bibitem [{\citenamefont {Kong}\ \emph {et~al.}(2015)\citenamefont {Kong},
  \citenamefont {Stark}, \citenamefont {Du}, \citenamefont {McGuinness},\ and\
  \citenamefont {Jelezko}}]{KSD+15}%
  \BibitemOpen
  \bibfield  {author} {\bibinfo {author} {\bibfnamefont {X.}~\bibnamefont
  {Kong}}, \bibinfo {author} {\bibfnamefont {A.}~\bibnamefont {Stark}},
  \bibinfo {author} {\bibfnamefont {J.}~\bibnamefont {Du}}, \bibinfo {author}
  {\bibfnamefont {L.~P.}\ \bibnamefont {McGuinness}}, \ and\ \bibinfo {author}
  {\bibfnamefont {F.}~\bibnamefont {Jelezko}},\ }\href@noop {} {\bibfield
  {journal} {\bibinfo  {journal} {Phys.\ Rev.\ Appl.}\ }\textbf {\bibinfo
  {volume} {4}},\ \bibinfo {pages} {024004} (\bibinfo {year}
  {2015})}\BibitemShut {NoStop}%
\bibitem [{\citenamefont {Perunicic}\ \emph {et~al.}(2016)\citenamefont
  {Perunicic}, \citenamefont {Hill}, \citenamefont {Hall},\ and\ \citenamefont
  {Hollenberg}}]{PHHH16}%
  \BibitemOpen
  \bibfield  {author} {\bibinfo {author} {\bibfnamefont {V.~S.}\ \bibnamefont
  {Perunicic}}, \bibinfo {author} {\bibfnamefont {C.~D.}\ \bibnamefont {Hill}},
  \bibinfo {author} {\bibfnamefont {L.~T.}\ \bibnamefont {Hall}}, \ and\
  \bibinfo {author} {\bibfnamefont {L.~C.~L.}\ \bibnamefont {Hollenberg}},\
  }\href@noop {} {\bibfield  {journal} {\bibinfo  {journal} {Nat.\ Commun.}\
  }\textbf {\bibinfo {volume} {7}},\ \bibinfo {pages} {12667} (\bibinfo {year}
  {2016})}\BibitemShut {NoStop}%
\bibitem [{\citenamefont {Kehayias}\ \emph {et~al.}(2017)\citenamefont
  {Kehayias}, \citenamefont {Jarmola}, \citenamefont {Mosavian}, \citenamefont
  {Fescenko}, \citenamefont {Benito}, \citenamefont {Laraoui}, \citenamefont
  {Smits}, \citenamefont {Bougas}, \citenamefont {Budker}, \citenamefont
  {Neumann}, \citenamefont {Brueck},\ and\ \citenamefont {Acosta}}]{KJM+17}%
  \BibitemOpen
  \bibfield  {author} {\bibinfo {author} {\bibfnamefont {P.}~\bibnamefont
  {Kehayias}}, \bibinfo {author} {\bibfnamefont {A.}~\bibnamefont {Jarmola}},
  \bibinfo {author} {\bibfnamefont {N.}~\bibnamefont {Mosavian}}, \bibinfo
  {author} {\bibfnamefont {I.}~\bibnamefont {Fescenko}}, \bibinfo {author}
  {\bibfnamefont {F.~M.}\ \bibnamefont {Benito}}, \bibinfo {author}
  {\bibfnamefont {A.}~\bibnamefont {Laraoui}}, \bibinfo {author} {\bibfnamefont
  {J.}~\bibnamefont {Smits}}, \bibinfo {author} {\bibfnamefont
  {L.}~\bibnamefont {Bougas}}, \bibinfo {author} {\bibfnamefont
  {D.}~\bibnamefont {Budker}}, \bibinfo {author} {\bibfnamefont
  {A.}~\bibnamefont {Neumann}}, \bibinfo {author} {\bibfnamefont {S.~R.~J.}\
  \bibnamefont {Brueck}}, \ and\ \bibinfo {author} {\bibfnamefont {V.~M.}\
  \bibnamefont {Acosta}},\ }\href@noop {} {\bibfield  {journal} {\bibinfo
  {journal} {Nat.\ Commun.}\ }\textbf {\bibinfo {volume} {8}},\ \bibinfo
  {pages} {188} (\bibinfo {year} {2017})}\BibitemShut {NoStop}%
\bibitem [{\citenamefont {Zhao}\ \emph {et~al.}(2011)\citenamefont {Zhao},
  \citenamefont {Hu}, \citenamefont {Ho}, \citenamefont {Wan},\ and\
  \citenamefont {Liu}}]{ZHH+11}%
  \BibitemOpen
  \bibfield  {author} {\bibinfo {author} {\bibfnamefont {N.}~\bibnamefont
  {Zhao}}, \bibinfo {author} {\bibfnamefont {J.-L.}\ \bibnamefont {Hu}},
  \bibinfo {author} {\bibfnamefont {S.-W.}\ \bibnamefont {Ho}}, \bibinfo
  {author} {\bibfnamefont {J.~T.~K.}\ \bibnamefont {Wan}}, \ and\ \bibinfo
  {author} {\bibfnamefont {R.~B.}\ \bibnamefont {Liu}},\ }\href@noop {}
  {\bibfield  {journal} {\bibinfo  {journal} {Nat.\ Nanotechnol.}\ }\textbf
  {\bibinfo {volume} {6}},\ \bibinfo {pages} {242} (\bibinfo {year}
  {2011})}\BibitemShut {NoStop}%
\bibitem [{\citenamefont {Laraoui}\ \emph {et~al.}(2011)\citenamefont
  {Laraoui}, \citenamefont {Hodges}, \citenamefont {Ryan},\ and\ \citenamefont
  {Meriles}}]{LHRM11}%
  \BibitemOpen
  \bibfield  {author} {\bibinfo {author} {\bibfnamefont {A.}~\bibnamefont
  {Laraoui}}, \bibinfo {author} {\bibfnamefont {J.~S.}\ \bibnamefont {Hodges}},
  \bibinfo {author} {\bibfnamefont {C.~A.}\ \bibnamefont {Ryan}}, \ and\
  \bibinfo {author} {\bibfnamefont {C.~A.}\ \bibnamefont {Meriles}},\
  }\href@noop {} {\bibfield  {journal} {\bibinfo  {journal} {Phys.\ Rev.\ B}\
  }\textbf {\bibinfo {volume} {84}},\ \bibinfo {pages} {104301} (\bibinfo
  {year} {2011})}\BibitemShut {NoStop}%
\bibitem [{\citenamefont {Kolkowitz}\ \emph {et~al.}(2012)\citenamefont
  {Kolkowitz}, \citenamefont {Unterreithmeier}, \citenamefont {Bennett},\ and\
  \citenamefont {Lukin}}]{KUBL12}%
  \BibitemOpen
  \bibfield  {author} {\bibinfo {author} {\bibfnamefont {S.}~\bibnamefont
  {Kolkowitz}}, \bibinfo {author} {\bibfnamefont {Q.~P.}\ \bibnamefont
  {Unterreithmeier}}, \bibinfo {author} {\bibfnamefont {S.~D.}\ \bibnamefont
  {Bennett}}, \ and\ \bibinfo {author} {\bibfnamefont {M.~D.}\ \bibnamefont
  {Lukin}},\ }\href@noop {} {\bibfield  {journal} {\bibinfo  {journal} {Phys.\
  Rev.\ Lett.}\ }\textbf {\bibinfo {volume} {109}},\ \bibinfo {pages} {137601}
  (\bibinfo {year} {2012})}\BibitemShut {NoStop}%
\bibitem [{\citenamefont {Taminiau}\ \emph {et~al.}(2012)\citenamefont
  {Taminiau}, \citenamefont {Wagenaar}, \citenamefont {van~der Sar},
  \citenamefont {Jelezko}, \citenamefont {Dobrovitski},\ and\ \citenamefont
  {Hanson}}]{TWS+12}%
  \BibitemOpen
  \bibfield  {author} {\bibinfo {author} {\bibfnamefont {T.~H.}\ \bibnamefont
  {Taminiau}}, \bibinfo {author} {\bibfnamefont {J.~J.~T.}\ \bibnamefont
  {Wagenaar}}, \bibinfo {author} {\bibfnamefont {T.}~\bibnamefont {van~der
  Sar}}, \bibinfo {author} {\bibfnamefont {F.}~\bibnamefont {Jelezko}},
  \bibinfo {author} {\bibfnamefont {V.~V.}\ \bibnamefont {Dobrovitski}}, \ and\
  \bibinfo {author} {\bibfnamefont {R.}~\bibnamefont {Hanson}},\ }\href@noop {}
  {\bibfield  {journal} {\bibinfo  {journal} {Phys.\ Rev.\ Lett.}\ }\textbf
  {\bibinfo {volume} {109}},\ \bibinfo {pages} {137602} (\bibinfo {year}
  {2012})}\BibitemShut {NoStop}%
\bibitem [{\citenamefont {Zhao}\ \emph {et~al.}(2012)\citenamefont {Zhao},
  \citenamefont {Honert}, \citenamefont {Schmid}, \citenamefont {Klas},
  \citenamefont {Isoya}, \citenamefont {Markham}, \citenamefont {Twitchen},
  \citenamefont {Jelezko}, \citenamefont {Liu}, \citenamefont {Fedder},\ and\
  \citenamefont {Wrachtrup}}]{ZHS+12}%
  \BibitemOpen
  \bibfield  {author} {\bibinfo {author} {\bibfnamefont {N.}~\bibnamefont
  {Zhao}}, \bibinfo {author} {\bibfnamefont {J.}~\bibnamefont {Honert}},
  \bibinfo {author} {\bibfnamefont {B.}~\bibnamefont {Schmid}}, \bibinfo
  {author} {\bibfnamefont {M.}~\bibnamefont {Klas}}, \bibinfo {author}
  {\bibfnamefont {J.}~\bibnamefont {Isoya}}, \bibinfo {author} {\bibfnamefont
  {M.}~\bibnamefont {Markham}}, \bibinfo {author} {\bibfnamefont
  {D.}~\bibnamefont {Twitchen}}, \bibinfo {author} {\bibfnamefont
  {F.}~\bibnamefont {Jelezko}}, \bibinfo {author} {\bibfnamefont {R.~B.}\
  \bibnamefont {Liu}}, \bibinfo {author} {\bibfnamefont {H.}~\bibnamefont
  {Fedder}}, \ and\ \bibinfo {author} {\bibfnamefont {J.}~\bibnamefont
  {Wrachtrup}},\ }\href@noop {} {\bibfield  {journal} {\bibinfo  {journal}
  {Nat.\ Nanotechnol.}\ }\textbf {\bibinfo {volume} {7}},\ \bibinfo {pages}
  {657} (\bibinfo {year} {2012})}\BibitemShut {NoStop}%
\bibitem [{\citenamefont {Laraoui}\ \emph {et~al.}(2013)\citenamefont
  {Laraoui}, \citenamefont {Dolde}, \citenamefont {Burk}, \citenamefont
  {Reinhard}, \citenamefont {Wrachtrup},\ and\ \citenamefont
  {Meriles}}]{LDB+13}%
  \BibitemOpen
  \bibfield  {author} {\bibinfo {author} {\bibfnamefont {A.}~\bibnamefont
  {Laraoui}}, \bibinfo {author} {\bibfnamefont {F.}~\bibnamefont {Dolde}},
  \bibinfo {author} {\bibfnamefont {C.}~\bibnamefont {Burk}}, \bibinfo {author}
  {\bibfnamefont {F.}~\bibnamefont {Reinhard}}, \bibinfo {author}
  {\bibfnamefont {J.}~\bibnamefont {Wrachtrup}}, \ and\ \bibinfo {author}
  {\bibfnamefont {C.~A.}\ \bibnamefont {Meriles}},\ }\href@noop {} {\bibfield
  {journal} {\bibinfo  {journal} {Nat.\ Commun.}\ }\textbf {\bibinfo {volume}
  {4}},\ \bibinfo {pages} {1651} (\bibinfo {year} {2013})}\BibitemShut
  {NoStop}%
\bibitem [{\citenamefont {Boss}\ \emph {et~al.}(2016)\citenamefont {Boss},
  \citenamefont {Chang}, \citenamefont {Armijo}, \citenamefont {Cujia},
  \citenamefont {Rosskopf}, \citenamefont {Maze},\ and\ \citenamefont
  {Degen}}]{BCA+16}%
  \BibitemOpen
  \bibfield  {author} {\bibinfo {author} {\bibfnamefont {J.~M.}\ \bibnamefont
  {Boss}}, \bibinfo {author} {\bibfnamefont {K.}~\bibnamefont {Chang}},
  \bibinfo {author} {\bibfnamefont {J.}~\bibnamefont {Armijo}}, \bibinfo
  {author} {\bibfnamefont {K.}~\bibnamefont {Cujia}}, \bibinfo {author}
  {\bibfnamefont {T.}~\bibnamefont {Rosskopf}}, \bibinfo {author}
  {\bibfnamefont {J.~R.}\ \bibnamefont {Maze}}, \ and\ \bibinfo {author}
  {\bibfnamefont {C.~L.}\ \bibnamefont {Degen}},\ }\href@noop {} {\bibfield
  {journal} {\bibinfo  {journal} {Phys.\ Rev.\ Lett.}\ }\textbf {\bibinfo
  {volume} {116}},\ \bibinfo {pages} {197601} (\bibinfo {year}
  {2016})}\BibitemShut {NoStop}%
\bibitem [{\citenamefont {Rosskopf}\ \emph {et~al.}(2017)\citenamefont
  {Rosskopf}, \citenamefont {Zopes}, \citenamefont {Boss},\ and\ \citenamefont
  {Degen}}]{RZB+17}%
  \BibitemOpen
  \bibfield  {author} {\bibinfo {author} {\bibfnamefont {T.}~\bibnamefont
  {Rosskopf}}, \bibinfo {author} {\bibfnamefont {J.}~\bibnamefont {Zopes}},
  \bibinfo {author} {\bibfnamefont {J.~M.}\ \bibnamefont {Boss}}, \ and\
  \bibinfo {author} {\bibfnamefont {C.~L.}\ \bibnamefont {Degen}},\ }\href@noop
  {} {\bibfield  {journal} {\bibinfo  {journal} {npj Quant.\ Inf.}\ }\textbf
  {\bibinfo {volume} {3}},\ \bibinfo {pages} {33} (\bibinfo {year}
  {2017})}\BibitemShut {NoStop}%
\bibitem [{\citenamefont {Laraoui}\ \emph {et~al.}(2015)\citenamefont
  {Laraoui}, \citenamefont {Pagliero},\ and\ \citenamefont {Meriles}}]{LPM15}%
  \BibitemOpen
  \bibfield  {author} {\bibinfo {author} {\bibfnamefont {A.}~\bibnamefont
  {Laraoui}}, \bibinfo {author} {\bibfnamefont {D.}~\bibnamefont {Pagliero}}, \
  and\ \bibinfo {author} {\bibfnamefont {C.~A.}\ \bibnamefont {Meriles}},\
  }\href@noop {} {\bibfield  {journal} {\bibinfo  {journal} {Phys.\ Rev.\ B}\
  }\textbf {\bibinfo {volume} {91}},\ \bibinfo {pages} {205410} (\bibinfo
  {year} {2015})}\BibitemShut {NoStop}%
\bibitem [{\citenamefont {Wang}\ \emph {et~al.}(2016)\citenamefont {Wang},
  \citenamefont {Haase}, \citenamefont {Casanova},\ and\ \citenamefont
  {Plenio}}]{WHCP16}%
  \BibitemOpen
  \bibfield  {author} {\bibinfo {author} {\bibfnamefont {Z.-Y.}\ \bibnamefont
  {Wang}}, \bibinfo {author} {\bibfnamefont {J.~F.}\ \bibnamefont {Haase}},
  \bibinfo {author} {\bibfnamefont {J.}~\bibnamefont {Casanova}}, \ and\
  \bibinfo {author} {\bibfnamefont {M.~B.}\ \bibnamefont {Plenio}},\
  }\href@noop {} {\bibfield  {journal} {\bibinfo  {journal} {Phys.\ Rev.\ B}\
  }\textbf {\bibinfo {volume} {93}},\ \bibinfo {pages} {174104} (\bibinfo
  {year} {2016})}\BibitemShut {NoStop}%
\bibitem [{\citenamefont {Wang}\ \emph {et~al.}(2017)\citenamefont {Wang},
  \citenamefont {Casanova},\ and\ \citenamefont {Plenio}}]{WCP17}%
  \BibitemOpen
  \bibfield  {author} {\bibinfo {author} {\bibfnamefont {Z.-Y.}\ \bibnamefont
  {Wang}}, \bibinfo {author} {\bibfnamefont {J.}~\bibnamefont {Casanova}}, \
  and\ \bibinfo {author} {\bibfnamefont {M.~B.}\ \bibnamefont {Plenio}},\
  }\href@noop {} {\bibfield  {journal} {\bibinfo  {journal} {Nat.\ Commun.}\
  }\textbf {\bibinfo {volume} {8}},\ \bibinfo {pages} {14660} (\bibinfo {year}
  {2017})}\BibitemShut {NoStop}%
\bibitem [{\citenamefont {Schwartz}\ \emph {et~al.}()\citenamefont {Schwartz},
  \citenamefont {Scheuer}, \citenamefont {Tratzmiller}, \citenamefont
  {M{\"u}ller}, \citenamefont {Chen}, \citenamefont {Dhand}, \citenamefont
  {Wang}, \citenamefont {M{\"u}ller}, \citenamefont {Naydenov}, \citenamefont
  {Jelezko},\ and\ \citenamefont {Plenio}}]{SST+17}%
  \BibitemOpen
  \bibfield  {author} {\bibinfo {author} {\bibfnamefont {I.}~\bibnamefont
  {Schwartz}}, \bibinfo {author} {\bibfnamefont {J.}~\bibnamefont {Scheuer}},
  \bibinfo {author} {\bibfnamefont {B.}~\bibnamefont {Tratzmiller}}, \bibinfo
  {author} {\bibfnamefont {S.}~\bibnamefont {M{\"u}ller}}, \bibinfo {author}
  {\bibfnamefont {Q.}~\bibnamefont {Chen}}, \bibinfo {author} {\bibfnamefont
  {I.}~\bibnamefont {Dhand}}, \bibinfo {author} {\bibfnamefont
  {Z.}~\bibnamefont {Wang}}, \bibinfo {author} {\bibfnamefont {C.}~\bibnamefont
  {M{\"u}ller}}, \bibinfo {author} {\bibfnamefont {B.}~\bibnamefont
  {Naydenov}}, \bibinfo {author} {\bibfnamefont {F.}~\bibnamefont {Jelezko}}, \
  and\ \bibinfo {author} {\bibfnamefont {M.~B.}\ \bibnamefont {Plenio}},\
  }\href@noop {} {}\bibinfo {note} {{a}rXiv:1710.01508
  (unpublished)}\BibitemShut {NoStop}%
\bibitem [{\citenamefont {Carr}\ and\ \citenamefont {Purcell}(1954)}]{CP54}%
  \BibitemOpen
  \bibfield  {author} {\bibinfo {author} {\bibfnamefont {H.~Y.}\ \bibnamefont
  {Carr}}\ and\ \bibinfo {author} {\bibfnamefont {E.~M.}\ \bibnamefont
  {Purcell}},\ }\href@noop {} {\bibfield  {journal} {\bibinfo  {journal}
  {Phys.\ Rev.}\ }\textbf {\bibinfo {volume} {94}},\ \bibinfo {pages} {630}
  (\bibinfo {year} {1954})}\BibitemShut {NoStop}%
\bibitem [{\citenamefont {Taminiau}\ \emph {et~al.}(2014)\citenamefont
  {Taminiau}, \citenamefont {Cramer}, \citenamefont {van~der Sar},
  \citenamefont {Dobrovitski},\ and\ \citenamefont {Hanson}}]{TCS+14}%
  \BibitemOpen
  \bibfield  {author} {\bibinfo {author} {\bibfnamefont {T.~H.}\ \bibnamefont
  {Taminiau}}, \bibinfo {author} {\bibfnamefont {J.}~\bibnamefont {Cramer}},
  \bibinfo {author} {\bibfnamefont {T.}~\bibnamefont {van~der Sar}}, \bibinfo
  {author} {\bibfnamefont {V.~V.}\ \bibnamefont {Dobrovitski}}, \ and\ \bibinfo
  {author} {\bibfnamefont {R.}~\bibnamefont {Hanson}},\ }\href@noop {}
  {\bibfield  {journal} {\bibinfo  {journal} {Nat.\ Nanotechnol.}\ }\textbf
  {\bibinfo {volume} {9}},\ \bibinfo {pages} {171} (\bibinfo {year}
  {2014})}\BibitemShut {NoStop}%
\bibitem [{SM()}]{SM}%
  \BibitemOpen
  \href@noop {} {}\bibinfo {note} {Supplemental Material}\BibitemShut {NoStop}%
\bibitem [{\citenamefont {Nizovtsev}\ \emph {et~al.}(2018)\citenamefont
  {Nizovtsev}, \citenamefont {Kilin}, \citenamefont {Pushkarchuk},
  \citenamefont {Pushkarchuk}, \citenamefont {Kuten}, \citenamefont {Zhikol},
  \citenamefont {Schmitt}, \citenamefont {Unden},\ and\ \citenamefont
  {Jelezko}}]{NKP+18}%
  \BibitemOpen
  \bibfield  {author} {\bibinfo {author} {\bibfnamefont {A.~P.}\ \bibnamefont
  {Nizovtsev}}, \bibinfo {author} {\bibfnamefont {S.~Y.}\ \bibnamefont
  {Kilin}}, \bibinfo {author} {\bibfnamefont {A.~L.}\ \bibnamefont
  {Pushkarchuk}}, \bibinfo {author} {\bibfnamefont {V.~A.}\ \bibnamefont
  {Pushkarchuk}}, \bibinfo {author} {\bibfnamefont {S.~A.}\ \bibnamefont
  {Kuten}}, \bibinfo {author} {\bibfnamefont {O.~A.}\ \bibnamefont {Zhikol}},
  \bibinfo {author} {\bibfnamefont {S.}~\bibnamefont {Schmitt}}, \bibinfo
  {author} {\bibfnamefont {T.}~\bibnamefont {Unden}}, \ and\ \bibinfo {author}
  {\bibfnamefont {F.}~\bibnamefont {Jelezko}},\ }\href@noop {} {\bibfield
  {journal} {\bibinfo  {journal} {New J.\ Phys.}\ }\textbf {\bibinfo {volume}
  {20}},\ \bibinfo {pages} {023022} (\bibinfo {year} {2018})}\BibitemShut
  {NoStop}%
\bibitem [{\citenamefont {Scheuer}\ \emph {et~al.}(2017)\citenamefont
  {Scheuer}, \citenamefont {Schwartz}, \citenamefont {M{\"u}llar},
  \citenamefont {Chen}, \citenamefont {Dhand}, \citenamefont {Plenio},
  \citenamefont {Naydenov},\ and\ \citenamefont {Jelezko}}]{SSM+17}%
  \BibitemOpen
  \bibfield  {author} {\bibinfo {author} {\bibfnamefont {J.}~\bibnamefont
  {Scheuer}}, \bibinfo {author} {\bibfnamefont {I.}~\bibnamefont {Schwartz}},
  \bibinfo {author} {\bibfnamefont {S.}~\bibnamefont {M{\"u}llar}}, \bibinfo
  {author} {\bibfnamefont {Q.}~\bibnamefont {Chen}}, \bibinfo {author}
  {\bibfnamefont {I.}~\bibnamefont {Dhand}}, \bibinfo {author} {\bibfnamefont
  {M.~B.}\ \bibnamefont {Plenio}}, \bibinfo {author} {\bibfnamefont
  {B.}~\bibnamefont {Naydenov}}, \ and\ \bibinfo {author} {\bibfnamefont
  {F.}~\bibnamefont {Jelezko}},\ }\href@noop {} {\bibfield  {journal} {\bibinfo
   {journal} {Phys.\ Rev.\ B}\ }\textbf {\bibinfo {volume} {96}},\ \bibinfo
  {pages} {174436} (\bibinfo {year} {2017})}\BibitemShut {NoStop}%
\bibitem [{\citenamefont {Wrachtrup}\ \emph {et~al.}(2013)\citenamefont
  {Wrachtrup}, \citenamefont {Jelezko}, \citenamefont {Grotz},\ and\
  \citenamefont {McGuinness}}]{WJGM13}%
  \BibitemOpen
  \bibfield  {author} {\bibinfo {author} {\bibfnamefont {J.}~\bibnamefont
  {Wrachtrup}}, \bibinfo {author} {\bibfnamefont {F.}~\bibnamefont {Jelezko}},
  \bibinfo {author} {\bibfnamefont {B.}~\bibnamefont {Grotz}}, \ and\ \bibinfo
  {author} {\bibfnamefont {L.}~\bibnamefont {McGuinness}},\ }\href@noop {}
  {\bibfield  {journal} {\bibinfo  {journal} {MRS Bull.}\ }\textbf {\bibinfo
  {volume} {38}},\ \bibinfo {pages} {149} (\bibinfo {year} {2013})}\BibitemShut
  {NoStop}%
\bibitem [{\citenamefont {Haeberlen}(1976)}]{H76}%
  \BibitemOpen
  \bibfield  {author} {\bibinfo {author} {\bibfnamefont {U.}~\bibnamefont
  {Haeberlen}},\ }\href@noop {} {\emph {\bibinfo {title} {High {R}esolution
  {NMR} in {S}olids {S}elective {A}veraging}}}\ (\bibinfo  {publisher}
  {Academic Press, New York},\ \bibinfo {year} {1976})\BibitemShut {NoStop}%
\bibitem [{\citenamefont {Schmitt}\ \emph {et~al.}(2017)\citenamefont
  {Schmitt}, \citenamefont {Gefen}, \citenamefont {St{\"u}rner}, \citenamefont
  {Unden}, \citenamefont {Wolff}, \citenamefont {M{\"u}ller}, \citenamefont
  {Scheuer}, \citenamefont {Naydenov}, \citenamefont {Markham}, \citenamefont
  {Pezzagna}, \citenamefont {Meijer}, \citenamefont {Schwarz}, \citenamefont
  {Plenio}, \citenamefont {Retzker}, \citenamefont {McGuinness},\ and\
  \citenamefont {Jelezko}}]{SGS+17}%
  \BibitemOpen
  \bibfield  {author} {\bibinfo {author} {\bibfnamefont {S.}~\bibnamefont
  {Schmitt}}, \bibinfo {author} {\bibfnamefont {T.}~\bibnamefont {Gefen}},
  \bibinfo {author} {\bibfnamefont {F.~M.}\ \bibnamefont {St{\"u}rner}},
  \bibinfo {author} {\bibfnamefont {T.}~\bibnamefont {Unden}}, \bibinfo
  {author} {\bibfnamefont {G.}~\bibnamefont {Wolff}}, \bibinfo {author}
  {\bibfnamefont {C.}~\bibnamefont {M{\"u}ller}}, \bibinfo {author}
  {\bibfnamefont {J.}~\bibnamefont {Scheuer}}, \bibinfo {author} {\bibfnamefont
  {B.}~\bibnamefont {Naydenov}}, \bibinfo {author} {\bibfnamefont
  {M.}~\bibnamefont {Markham}}, \bibinfo {author} {\bibfnamefont
  {S.}~\bibnamefont {Pezzagna}}, \bibinfo {author} {\bibfnamefont
  {J.}~\bibnamefont {Meijer}}, \bibinfo {author} {\bibfnamefont
  {I.}~\bibnamefont {Schwarz}}, \bibinfo {author} {\bibfnamefont
  {M.}~\bibnamefont {Plenio}}, \bibinfo {author} {\bibfnamefont
  {A.}~\bibnamefont {Retzker}}, \bibinfo {author} {\bibfnamefont {L.~P.}\
  \bibnamefont {McGuinness}}, \ and\ \bibinfo {author} {\bibfnamefont
  {F.}~\bibnamefont {Jelezko}},\ }\href@noop {} {\bibfield  {journal} {\bibinfo
   {journal} {Science}\ }\textbf {\bibinfo {volume} {356}},\ \bibinfo {pages}
  {832} (\bibinfo {year} {2017})}\BibitemShut {NoStop}%
\bibitem [{\citenamefont {Boss}\ \emph {et~al.}(2017)\citenamefont {Boss},
  \citenamefont {Cujia}, \citenamefont {Zopes},\ and\ \citenamefont
  {Degen}}]{BCZD17}%
  \BibitemOpen
  \bibfield  {author} {\bibinfo {author} {\bibfnamefont {J.~M.}\ \bibnamefont
  {Boss}}, \bibinfo {author} {\bibfnamefont {K.~S.}\ \bibnamefont {Cujia}},
  \bibinfo {author} {\bibfnamefont {J.}~\bibnamefont {Zopes}}, \ and\ \bibinfo
  {author} {\bibfnamefont {C.~L.}\ \bibnamefont {Degen}},\ }\href@noop {}
  {\bibfield  {journal} {\bibinfo  {journal} {Science}\ }\textbf {\bibinfo
  {volume} {356}},\ \bibinfo {pages} {837} (\bibinfo {year}
  {2017})}\BibitemShut {NoStop}%
\bibitem [{\citenamefont {Zopes}\ \emph {et~al.}()\citenamefont {Zopes},
  \citenamefont {Cujia}, \citenamefont {Sasaki}, \citenamefont {Boss},
  \citenamefont {Itoh},\ and\ \citenamefont {Degen}}]{ZCS+18}%
  \BibitemOpen
  \bibfield  {author} {\bibinfo {author} {\bibfnamefont {J.}~\bibnamefont
  {Zopes}}, \bibinfo {author} {\bibfnamefont {K.~S.}\ \bibnamefont {Cujia}},
  \bibinfo {author} {\bibfnamefont {K.}~\bibnamefont {Sasaki}}, \bibinfo
  {author} {\bibfnamefont {J.~M.}\ \bibnamefont {Boss}}, \bibinfo {author}
  {\bibfnamefont {K.~M.}\ \bibnamefont {Itoh}}, \ and\ \bibinfo {author}
  {\bibfnamefont {C.~L.}\ \bibnamefont {Degen}},\ }\href@noop {} {}\bibinfo
  {note} {{a}rXiv:1806.04883 (unpublished)}\BibitemShut {NoStop}%
\bibitem[S1]{SGS+17s}
S. Schmitt, T. Gefen, F. M. St{\"u}rner, T. Unden, G. Wolff, C. M{\"u}ller, J. Scheuer, B. Naydenov, M. Markham, S. Pezzagna, J. Meijer, I. Schwarz, M. Plenio, A. Retzker, L. P. McGuinness, and F. Jelezko,
Science {\bf356}, 832 (2017).
\bibitem[S2]{KF95s}
E. Kup{\v{c}}e and R. Freeman, J.\ Magn.\ Reson. {\bf115}, 273 (1995).
\bibitem[S3]{SSB+16s}
P. E. Spindler, P. Sch{\"o}ps, A. M. Bowen, B. Endeward, and T. F. Prisner,
eMagRes, {\bf5}, 1477 (2016).
\bibitem[S4]{TWS+12s}
T. H. Taminiau, J. J. T. Wagenaar, T. van der Sar, F. Jelezko, V. V. Dobrovitski, and R. Hanson,
Phys. Rev. Lett. {\bf109}, 137602 (2012).
\bibitem[S5]{BCA+16s}
J. M. Boss, K. Chang, J. Armijo, K. Cujia, T. Rosskopf, J. R. Maze, and C. L. Degen,
Phys. Rev. Lett. {\bf116}, 197601 (2016).
\bibitem[S6]{AS18s}
E. Abe and K. Sasaki,
J. Appl. Phys. {\bf123}, 161101 (2018).
\bibitem[S7]{SST+17s}
I. Schwartz, J. Scheuer, B. Tratzmiller, S. M{\"u}ller, Q. Chen, I. Dhand, Z. Wang, C. M{\"u}ller, B. Naydenov, F. Jelezko, and M. B. Plenio,
arXiv:1710.01508 (unpublished).
\bibitem[S8]{HDSW88s}
A. Henstra, P. Dirksen, J. Schmidt, and W. T. Wenckebach,
J. Magn. Reson. {\bf77}, 389 (1988).
\bibitem[S9]{SSC+16s}
J. Scheuer, I. Schwartz, Q. Chen, D. Schulze-S{\''u}nninghausen, P. Carl, P. H{\''o}fer, A. Retzker, H. Sumiya, J. Isoya, B. Luy, M. B. Plenio, B. Naydenov, and F. Jelezko,
New J. Phys. {\bf18}, 013040 (2016).
\bibitem[S10]{GBL+18s}
D. R. Glenn, D. B. Bucher, J. Lee, M. D. Lukin, H. Park, and R. L. Walsworth,
Nature {\bf555}, 351 (2018).
\bibitem[S11]{BCZD17s}
J. M. Boss, K. S. Cujia, J. Zopes, and C. L. Degen,
Science {\bf356}, 837 (2017).\end{thebibliography}
%
\end{document}